\documentclass[number,reqno]{elsarticle}
\usepackage{graphicx}
\usepackage{epstopdf}
\usepackage{subfigure}
\usepackage{amsmath}
\usepackage{amsthm}
\usepackage{amsfonts}
\usepackage{amssymb}
\usepackage{float}

\begin{document}



\title{A space-time smooth artificial viscosity method for nonlinear conservation laws }

\begin{frontmatter}

\author{J. Reisner}
\ead{reisner@lanl.gov}
\address{Los Alamos National Lab, XCP-4 MSF605, Los Alamos, NM 87544}

\author{J.  Serencsa\corref{cor1}}
\ead{jserencs@math.ucsd.edu}
\address{Department of Mathematics, UC Davis, One Shields Ave., Davis, CA 95616}

\author{S. Shkoller\corref{cor2}}
\ead{shkoller@math.ucdavis.edu}
\address{Department of Mathematics, UC Davis, One Shields Ave., Davis, CA 95616}

\begin{abstract}
We introduce a new methodology for adding localized, space-time smooth, artificial viscosity to nonlinear systems of conservation laws
which propagate shock waves, rarefactions, and contact discontinuities, which we call the $C$-method.  We shall focus our attention on the compressible
Euler equations in one space dimension.   The novel feature of our approach involves the coupling
of a linear scalar reaction-diffusion equation to our system of conservation laws, whose solution $C(x,t)$ is the coefficient to an
additional (and artificial) term added to the  flux, which determines  the location, localization, and strength of the artificial viscosity.
Near shock discontinuities, $C(x,t)$ is large and localized, and transitions smoothly in space-time  to zero away from discontinuities.  
Our approach is a provably convergent, spacetime-regularized variant of the original idea of Richtmeyer and Von Neumann, and
is provided at the level of the PDE, thus allowing a host of numerical discretization schemes to be employed.

We demonstrate the effectiveness of the $C$-method with three  different  numerical implementations and apply these to
a collection of classical problems:   the Sod shock-tube, the Osher-Shu shock-tube,
the Woodward-Colella blast wave and the Leblanc shock-tube.   First, we use a classical continuous finite-element implementation
using second-order discretization in both space and time , FEM-C. Second,  we use a simplified
WENO scheme within our $C$-method framework, WENO-C.  Third, we use WENO with the Lax-Friedrichs flux together with
the $C$-equation, and call this WENO-LF-C.  All three schemes yield  higher-order  discretization strategies, which provide sharp shock resolution
with minimal overshoot and noise, and compare well with higher-order WENO schemes that employ  approximate Riemann solvers, outperforming
them for the difficult Leblanc shock tube experiment.
\end{abstract}

\end{frontmatter}


\section{Introduction}  \label{sec:intro}

\subsection{Smoothing conservation laws}
The initial-value problem for a general nonlinear
system of conservation laws can be written as an evolution equation,
\begin{equation}\label{scl}
\partial_t U (x,t) + \operatorname{div} F(U(x,t)) = 0 \text{ with } U|_{t=0} =U_0\,,
\end{equation} 
for an $m$-vector $U$ defined on ($D$$+$$1$)-dimensional space-time.   
Such partial differential equations (PDE) are both ubiquitous and fundamental in
science and engineering, and include the compressible Euler equations of gas dynamics, the magneto-hydrodynamic  (MHD) equations modeling
ionized plasma,  the elasticity equations of solid mechanics,  and numerous related physical systems which 
possess complicated nonlinear wave interactions.

It is well known that solutions of (\ref{scl}) can develop finite-time shocks, even when the initial data is smooth, in 
which case,  discontinuities of $U$ are 
 propagated according to the so-called Rankine-Hugoniot conditions (see Section \ref{subsec:compEuler} below).   
It is important to develop stable and robust numerical algorithms which can approximate weak shock-wave solutions.
Even
in one-space dimension, nonlinear wave interaction such as two shock waves colliding,  is a difficult problem when considering
accuracy, stability and monotonicity.   The challenge is maintaining higher-order accuracy away from the shock while approximating
the discontinuity in an order-$ \Delta x$ smooth transition region where $ \Delta x$ denotes the spatial grid size.
 
 As we describe below, a variety of
clever discretization schemes have been developed and employed, particularly in one-space dimension, to approximate discontinuous solution profiles
in an essentially non-oscillatory (ENO) fashion. 
  These include, but are not limited to,  total variation diminishing (TVD) schemes, flux-corrected transport (FCT) schemes, 
weighted essentially non-oscillatory (WENO) schemes, discontinuous Galerkin methods, artificial diffusion methods,  exact and approximate Riemann solvers, and a host of
variants and combinations of these techniques. 

We develop a  robust parabolic-type regularization of (\ref{scl}), which we refer to as the $C$-method, which couples a modified set of  $m$ equations for $U$ with an
 additional linear scalar reaction-diffusion
equation for a new scalar field $C(x,t)$.   Thus, instead of (\ref{scl}) we consider
a system of $m$$+$$1$  equations, which use the solution $C(x,t)$ as a coefficient in a carefully chosen modification of the
 flux.  As we describe in detail below, the solution $C(x,t)$ is highly localized in
regions of discontinuity, and transitions smoothly (in both $x$ and $t$) to zero in regions wherein the solution is smooth.   Further, as $\Delta x \to 0$, we recover the original 
hyperbolic nonlinear system of conservation laws (\ref{scl}).

\subsection{Numerical discretization}
In the case of  1-D gas dynamics,  the construction of non-oscillatory, higher-order,  numerical algorithms such as ENO by Harten, Engquist,  Osher \& Chakravarthy \cite{Harten1987231} and Shu \& Osher \cite{Shu1988439}, \cite{Shu198932};  WENO by Liu, Osher, \& Chan \cite{LiOsCh1994} and
Jiang \& Shu \cite{Jiang1996202}; MUSCL by Van Leer \cite{VanLeer1979101},  Colella \cite{Colella1985104}, and  Huynh \cite{Huynh19951565}; or PPM by Colella \& Woodward \cite{Colella1984174}  requires 
carefully chosen {\it reconstruction} and {\it numerical flux}.

Such numerical methods evolve cell-averaged quantities;  to calculate an accurate approximation of the flux at cell-interfaces, these schemes  reconstruct $k$th-order ($k\ge 2$) polynomial
approximations of the solution (and hence the flux) from the computed cell-averages, and thus provide $k$th-order accuracy away from discontinuities.   See, for example, the convergence plots
of Greenough \& Rider \cite{Greenough2004259} and  Liska \& Wendroff \cite{Liska2003995}.  Given a polynomial representation of the solution, a strategy is chosen to compute the
most accurate cell-interface flux, and this is achieved by a variety of algorithms.
Centered numerical fluxes, such as Lax-Friedrichs, add dissipation as a mechanism to preserve stability and monotonicity. On the other hand, {\it characteristic-type} upwinding based upon exact (Godunov) or approximate (Roe, Osher, HLL, HLLC) Riemann solvers,
 which preserve monotonicity without adding too much dissipation,  tend to be rather complex and PDE-specific;  moreover,  for strong shocks, other techniques may be
 required  to dampen post-shock oscillations or to yield entropy-satisfying approximations (see Quirk \cite{Quirk1994555}).
 Again, we refer the reader to the papers  \cite{Greenough2004259}, \cite{Liska2003995} or Colella \& Woodward \cite{Colella1984115} for a thorough overview, as well as a comparison of the effectiveness of a variety of competitive schemes.
 
 Majda \& Osher \cite{Majda1977671} have shown  that {\em any} numerical scheme is  {\em at best}, first-order accurate in the presence of shocks or discontinuities.  The use of
 higher-order numerical schemes is, nevertheless, imperative for the  elimination of error-terms in the Taylor expansion (in mesh-size)  and the  subsequent limiting of truncation error. 
 Moreover, higher-order schemes tend to be less dissipative than there lower-order counterparts, as discussed by Greenough \& Rider \cite{Greenough2004259}; therein,
 a comparison between  a $2$nd-order PLMDE scheme and a  $5$th-order WENO scheme demonstrates the improved resolution of  intricate  fine structure  afforded by
 $5$th-order WENO,   while simultaneously providing far less clipping of local extrema than PLMDE.

In multi-D, similar tools are required to obtain non-oscillatory numerical schemes, but the multi-dimensional analogues  to those described above are generally limited by mesh considerations. For structured grids 
(such as products of uniform 1-D grids), dimensional splitting is commonly used, decomposing the problem into a sequence of 1-D problems. This technique is quite successful, but stringent mesh requirements prohibits its use on complex domains. Moreover, applications to PDE outside of variants of the Euler equations may be somewhat limited. For further discussion of the limitations of dimensional splitting, we refer the reader to Crandall \& Majda \cite{Crandall80}, and  Jiang \& Tadmor \cite{Jiang98}. For unstructured grids, dimensional splitting is not available and alternative approaches  must be employed, 
necessitated by the lack of multi-D Riemann solvers. WENO schemes on unstructured  triangular grids have been developed in Hu \& Shu \cite{Hu199997}, but using simplified methods, which employ reduced characteristic decompositions,  can lead   to a loss of   monotonicity and stability.

Algorithms that explicitly introduce diffusion provide a simple way to stabilize higher-order numerical schemes and subsequently remove non-physical oscillations near shocks.
In the mathematical analysis of conservation laws (and in the truncation error of certain discretization schemes), the simplest parabolic-regularization is by the
addition of a 
 uniform linear  viscosity.   Choosing a constant $\beta > 0$, which depends upon mesh-size $\Delta x$ and sometimes velocity or wave-speed,  and adding  
\begin{equation}\label{ss5}
\beta (\Delta x)  \partial^2_{x} U(x,t)
\end{equation} 
to the right hand side of \eqref{scl}
provides a uniformly parabolic regularization of the hyperbolic conservation laws, and its discrete implementation smears sharp discontinuities across  $ O(\Delta x)$-regions and
thus adds stabilization, but unfortunately,  at the cost of accuracy. 
 With the addition of uniform linear viscosity,  shocks and discontinuities are captured in a non-oscillatory fashion, but  the transition region form left to right state,  which approximates the  discontinuity, tends to grow over time. Moreover, since viscosity is applied uniformly over the  {\em entire} domain $\mathcal{I} $ , the benefits of a higher-order scheme (away from the discontinuity) may be  lost,  and the accuracy  may often reduce to merely first-order. 
In practical implantation in a numerical scheme,  the use of viscosity should be localized in regions of shock (so as to stabilize the scheme),  limited at contact discontinuities (to avoid over-smearing the sharp transition), and very small in smooth regions away from discontinuities.
Achieving these requirements allows higher-order approximation of smooth flow and sharp, non-oscillatory, resolution of shocks and discontinuities.
Naturally, this necessitates that the amount of added viscosity be a function of the solution.

The pioneering papers of  Richtmyer \cite{Richtmyer1948},  Von Neumann \& Richtmyer \cite{VonNeumann1950232}, Lax \& Wendroff \cite{Lax1960217}, and Lapidus  \cite{Lapidus1967154} suggest  the introduction of nonlinear  artificial viscosity to equations \eqref{scl} in a form similar to the following expression:
\begin{equation}\label{ss4}
\beta (\Delta x)^2 \partial_x \left ( \left | \partial_x u(x,t) \right | \partial_x U(x,t) \right ).
\end{equation} 
We refer the reader to the classical papers of Gentry, Martin, \& Daly \cite{GeMaDa1966} and Harlow \& Amsden \cite{HaAm1971} for an interesting discussion on artificial viscosity.  Specifically, 
Gentry, Martin, \& Daly \cite{GeMaDa1966} define  the nonlinear viscosity  of the type (\ref{ss4}) to be artificial viscosity, and show that the  linear viscosity (\ref{ss5}),  scaled by the magnitude of local velocity,  arises
as  truncation error (in finite-difference approximations).   The latter is responsible for stabilizing the {\it transport} of sound waves, while (\ref{ss4}) stabilizes the {\it steepening} of sound waves.\footnote{We are
indebted to the anonymous referee for clarifying this point for us.}

We are primarily concerned with the steepening of sound waves, and shall term artificial viscosity of the type (\ref{ss4}) as {\it classical artificial viscosity}.
Formally, the use of (\ref{ss4})  produces the required amount of viscosity near shocks but allows for second-order accuracy in smooth regions.
On the other hand, the diffusion coefficient $|\partial_x u(x,t)|$ is precisely  the quantity which loses regularity (or smoothness) near shock discontinuities. Also, the constant $\beta$ must be larger than one to control numerical oscillations behind
the shock wave, which in turn overly diffuses the waves and produces incorrect wave speeds.

Alternative procedures have been proposed. For streamline upwind Petrov-Galerkin schemes (SUPG), Hughes \& Mallet \cite{Hughes1986329} and Shakib, Hughes, \& Johan \cite{Shakib1991141} use residual-based artificial viscosity. Guermond \& Pasquetti \cite{Guermond2008801} present a similar, entropy-residual-based scheme for use in spectral methods. Persson \& Peraire \cite{Persson2006112} develop a method based upon decay of local interpolating polynomials for discontinuous Galerkin schemes.  Later, Barter \& Darmofal \cite{Barter20101810} use a reaction-diffusion equation to provide a regularized variant of this approach.

Our approach is similar to \cite{Barter20101810} in that it uses a reaction-diffusion equation to calculate a smooth distribution of artificial viscosity. Instead of regularizing a DG-based noise-indicator that allows for the growth of
viscosity near shocks, we regularize the classical artificial viscosity of \cite{Lapidus1967154}, using a gradient based approach for this source term. This approach yields both a discretization- and PDE-independent methodology which can be generalized to multiple dimensions by regularizing a similar viscosity to that in L{\"o}hner, Morgan, \& Peraire \cite{Lohner1985141}.

In 1-D, our approach proves to be a simple way of circumventing the need for characteristic or other {\it a priori} information of the exact solution to remove oscillations in higher-order schemes. 
 Due to the simple and discretization-independent nature of our method, we expect our methodology to be useful for a wide range of applications.

\subsection{Outline of the paper}

In Section \ref{sec:Cmethod}, we introduce the $C$-method for the compressible Euler equations in one space dimension.  We show that the
$C$-method is Galilean invariant and  that solutions of the $C$-method  converge to the unique weak solutions of the Euler equations in the limit of zero mesh size.   We also
show the relative smoothness of our new viscosity coefficient with respect to the classical artificial viscosity of Richtmyer and Von Neumann, and we demonstrate the ability
of the $C$-method to remove downstream oscillation in slowly moving shocks.

In Section \ref{sec:numSchemes}, we give a brief outline of the  numerical schemes whose solutions are used in this paper.
First, we outline a second-order, continuous Galerkin finite-element method. Second, we outline a simple WENO-based finite-volume scheme, only upwinding via the sign of the velocity (no Riemann-solvers or characteristic decompositions in primitive variables). The resulting schemes applied to the $C$-method are referred to as FEM-C and WENO-C, respectively. Third, we outline the central-finite-difference scheme of Nessyahu and Tadmor (NT), a simple scheme, easily generalizable to multi-D \cite{Nessyahu1990408}. Like our FEM-C scheme, the NT-scheme is at best, second-order, and does not require specialized techniques for upwinding. Fourth, we outline a  Godunov-type characteristic decomposition-based WENO scheme (WENO-G) developed by Rider, Greenough  \& Kamm \cite{rideretal} which utilizes a variant of a Godunov/Riemann-solver as upwinding, providing a very competitive scheme for modeling the collision of very strong shocks.

In Section \ref{sec:sodshock}, we consider the classical shock-tube problem of Sod.  With the Sod shock problem, we apply our FEM-C scheme and compare with the classical viscosity approach. We then compare the FEM-C scheme with the two standalone methods, NT and WENO-G.

In Section \ref{sec:osherShu}, we consider the moderately difficult problem of Osher-Shu, modeling the interaction of a mild shock with an entropy wave. We compare FEM-C to NT and WENO-G in which the differences are more significant than in the Sod-shock comparisons. We show that  WENO-C compares well with WENO-G;  on the other hand, the  simple WENO scheme without the $C$-method and
without the Gudonov-based characteristic solver also does well in modeling the Osher-Shu test case.

In Section \ref{sec:woodwardColella},  we consider the numerically challenging Woodward-Colella blast wave simulation, which models  the collision of two strong 
interacting shock fronts. Though the FEM-C scheme performs  better than NT, both second-order schemes are somewhat out-performed by  the higher-order
 WENO-G method (with characteristic solver).
On the other hand,  WENO-C compares well with   WENO-G, having slightly less damped amplitudes with the same shock resolution.

Finally, in Section \ref{sec:leblanc}, we consider the Leblanc shock-tube, an extremely difficult test case consisting of a very strong shock. 
For this problem, devise two strategies to demonstrate the use of the $C$-method.  In the first strategy, we use our simplified WENO-C scheme
with a right-hand side term for the energy equation that relies on a second $C$-equation which smooths gradients of $E/\rho$.   We obtain
 an excellent approximation  of the notoriously difficult
contact discontinuity for internal energy, while maintaining an accurate shock speed;  simultaneously,  we avoid generating large overshoots  at the contact discontinuity,
which would indeed occur without the use of the $C$-method.  For our second strategy, we show that WENO with the  Lax-Friedrichs flux can be significantly
improved with the addition of the $C$-method.  We call this algorithm WENO-LF-C, and show that by using just one $C$-equation (as we have for all of the other
test cases), we can sharply resolve the contact discontinuity for the internal energy, with accurate wave speed, and without overshoots.

\section{The $C$-method}
\label{sec:Cmethod}

We begin with a description of the 1-D compressible Euler equations, written as a 3x3 system of conservation laws.    We then explain our
parabolic regularization, which we call the $C$-method.

\subsection{Compressible Euler equations} 
\label{subsec:compEuler}

The  compressible Euler equations set on a 1-D space domain $ \mathcal{I} \subset \mathbb{R}  $, and a time interval $[0,T]$  
are written in vector-form as the following coupled system of nonlinear conservation laws:
 \begin{subequations}
\label{subeq:consLaw}
\begin{alignat}{2}
\partial_t {\bf u}(x,t)+ \partial_x {\bf F}({\bf u}(x,t)) & = 0, && \ \ \  x\in \mathcal{I} \,,   t > 0, \label{eqn:consLawEvolution} \\
{\bf u}(x,0) & = {\bf u}_0(x), && \ \ \ x\in \mathcal{I} \,,   t = 0, \label{eqn:consLawIC}
\end{alignat}
\end{subequations}
where the $3$-vector ${\bf u}(x,t)$ and {\it flux function} ${\bf F}({\bf u}(x,t))$ are defined, respectively, as 
$$
{\bf u} = \left ( \begin{array}{c} \rho \\ m \\ E \end{array} \right ) \quad \text{and} \quad {\bf F}({\bf u}) = \left ( \begin{array}{c} m \\ \frac{m^2}{\rho} + p \\ \frac{m}{\rho} \left ( E + p \right ) \end{array} \right )\,,
$$
and 
\[
{\bf u}_0(x) = \left (\begin{array}{c} \rho_0(x) \\  m_0(x) \\ E_0(x) \end{array}\right ) 
\]
denotes the initial data for the problem.   The variables $\rho, \ m$, and $E$ denote the {\it density},  {\it momentum}, and {\it energy density} of a compressible gas,
while $p= \mathcal{H}  (\rho, m, E)$ denotes the {\it pressure} function.   It is necessary to choose an equation-of-state $ \mathcal{H}(\rho, m, E)$, and we
use the  ideal gas law, for which
\begin{equation}\label{eos}
p = (\gamma - 1) \left ( E - \frac{m^2}{2 \rho} \right )\,,
\end{equation} 
where $ \gamma $ denotes  the adiabatic constant.    The equations (\ref{subeq:consLaw}) are indeed conservation laws, as they represent the conservation
of mass, momentum, and energy in the evolution of a compressible gas.    The velocity field $u(x,t)$ is obtained from momentum and density via the identity
$$
u = \frac{m}{\rho}\,.
$$
Inverting the relation (\ref{eos}), we see that the energy density $E$ is a sum of kinetic and potential energy density functions:
$$
E= \underbrace{\frac{\rho\, u^2}{2}}_{\text{kinetic}} + \underbrace{\frac{p}{\gamma -1}}_{\text{potential}} \,.
$$
The gradient (or Jacobian) of the flux vector ${\bf F}({\bf u})$ is given  by
$$
D{\bf F}({\bf u}) =
\left[
\begin{array}{ccc}
0 & 1 & 0 \\
\frac{(\gamma -3) m^2}{2\rho^2} & \frac{(3-\gamma) m}{\rho} & \gamma -1 \\
- \gamma \frac{Em}{\rho^2} + (\gamma -1)  \frac{m^3}{\rho^3} &
          \frac{ \gamma E}{\rho} + (1- \gamma ) \frac{ 3 m^2}{ 2\rho^2} & \frac{ \gamma m}{\rho}
\end{array}\right]
$$
with eigenvalues
\begin{subequations}
\label{subeq:maxWaveSpeed}
\begin{equation}
  \lambda_1 = u + c\,, \ \lambda_2 = u\,,  \ \lambda_3 = u - c \,,
\end{equation}
\end{subequations}
where $c$ denotes the sound speed
(see, for example, Toro \cite{Toro2009}).   These eigenvalues determine the wave speeds.

The behavior of the various wave patterns is greatly influenced by the speed of propagation; as such, we define
the maximum {\it wave speed} to be

\begin{equation*}
[S({\bf u})](t) = \max_{i =1,2,3} \max_{x \in \mathcal{I}} \left \{ |\lambda_i(x,t)| \right \} \,.
\tag{\ref{subeq:maxWaveSpeed}b}
\end{equation*}
as a function of $t$.

\def\u{{\bf u}} 

We are interested in solutions with shock waves and contact discontinuities.      The  Rankine-Hugoniot  (R-H) conditions
determine the speed $s$ of the moving shock discontinuity, as well as the speed of a contact discontinuity.  For a shock wave
discontinuity, the R-H condition can be stated
$$
F(\u _l) - F(\u _r) = s ( \u_l - \u_r)
$$
where the subscript $l$ denotes the state to the left of the discontinuity, and the subscript $r$ denotes the state to the right of
the discontinuity.  In general, 
the following three jump conditions must hold:

\begin{align*}
m_l - m_r & = s(\rho_l - \rho_r) \\
\left(\frac{(3- \gamma ) m_l^2}{ 2 \rho_l^2} + (\gamma -1) E_l \right) - 
\left(\frac{(3- \gamma ) m_r^2}{ 2 \rho_r^2} + (\gamma -1) E_r \right) & = s (m_l - m_r) \\
\left(\gamma \frac{ E_l m_l}{\rho_l} - \frac{ \gamma -1}{2} \frac{m_l^3}{\rho_l^2}\right) &= s(E_l -E_r) \,.
\end{align*} 

There can be non-uniqueness for weak solutions that have jump discontinuities, unless entropy conditions are satisfied
(see the discussion in Section \ref{sec::entropy}).   So-called viscosity solutions  $\u_{vis}$ are known to satisfy the entropy condition
(and are hence unique) and are defined as the limit
as $ \epsilon \to 0$ of a sequence of solutions $\u^ \epsilon$ to the following parabolic equation:
\begin{subequations}
\label{subeq:strongVanishingViscosity}
\begin{alignat}{2}
\partial_t {\bf u}^\epsilon + \partial_x {\bf F}({\bf u}^ \epsilon ) & =  \epsilon  \partial_{xx} {\bf u}^ \epsilon  , &&  \quad t > 0, \label{eqn:strongVanishingViscosity1} \\
{\bf u}^ \epsilon  & = {\bf u}_0, && \quad t = 0\,. \label{eqn:strongVanishingViscosity2}
\end{alignat}
\end{subequations}
In the isentropic setting,  for bounded initial data $\u_0$ with bounded variation, solutions $\u^ \epsilon$ converge to the unique entropy solution $\u_{vis}$
 of  (\ref{subeq:consLaw})  as $ \epsilon \to 0$ (see DiPerna \cite{DiPerna1983} and Lions, Perthame, \& Souganidis \cite{LiPeSo1996}).  For  non-isentropic dynamics, the same result holds if the initial data has small total variation
(see Bianchini \& Bressan \cite{BiBr2005}).
Moreover, if the initial data $\u_0$ is regularized,
then solutions to (\ref{subeq:strongVanishingViscosity}) are smooth in both space and time, and the discontinuity is approximated
by a smooth function, transitioning from the left-state to the right-state over an interval whose length is $O( \epsilon )$.

Some of the classical finite-differencing schemes, such as the Lax-Friedrichs  discretization, is dissipative to second-order and effectively behaves as a 
discrete version of (\ref{subeq:strongVanishingViscosity}). The  uniform
nature of such diffusion does not distinguish between discontinuous and smooth flow regimes, and thus adds unnecessary dissipation in
regions of the wave profile which do not require any numerical stabilization.    Such uniform dissipation 
contributes to a non-phyiscal damping of entropy waves, over-diffusion and smearing of contact discontinuities, and changes the
wave speeds. Ultimately, uniform artificial viscosity is not ideal; rather, artificial viscosity should
 be added in a localized and smooth manner.
 
\subsection{Classical artificial viscosity}
\label{subsec:classicalViscosity}

The idea of adding localized artificial viscosity to capture discontinuous solution profiles in numerical simulations dates back to Richtmyer  \cite{Richtmyer1948},
Von Neumann \& Richtmyer \cite{VonNeumann1950232},  Lax \& Wendroff \cite{Lax1960217},  Lapidus \cite{Lapidus1967154} and a host of other reseachers.
The idea behind {\it classical artificial viscosity} is to  refine the uniform viscosity on the right-hand side of equation  \eqref{eqn:strongVanishingViscosity1} with
\begin{equation}
\partial_t {\bf u}^\epsilon + \partial_x {\bf F} ( {\bf u}^\epsilon) =  \beta \epsilon^2 \partial_x (|\partial_x u^\epsilon|  \partial_x {\bf u}^\epsilon),  \quad t  > 0\,, \label{eqn:classicalViscosityEuler}
\end{equation}
for a suitably chosen constant $\beta > 0$, which may depend upon the numerical discretization scheme.

When the velocity $u$ exhibits a jump discontinuity (i.e., at a shock),  the quantity $|\partial_x u^\epsilon|$ is $O(\frac{1}{\epsilon})$; however, away from shocks, where the velocity is smooth, $|\partial_x u^\epsilon|$ remains uniformly bounded in $\epsilon$, and in such smooth regions,  \eqref{eqn:classicalViscosityEuler} adds significantly less viscosity than \eqref{eqn:strongVanishingViscosity1}.    On the other hand, as we shall demonstrate in Figure \ref{fig:spacetimesmoothing}, the use of $|\partial _x u^ \epsilon |$ as a coefficient in the smoothing operator, can lead to
spurious oscillations in the solution, caused by the lack of regularity in the quantity $|\partial_x u^ \epsilon |$.  

Formally, the use of the localizing coefficient $|\partial_xu^ \epsilon |$ corrects for the over-dissipation of the uniform viscosity in  \eqref{subeq:strongVanishingViscosity},
and a variety of  numerical methods  have employed some variant of this idea, achieving methods that are nominally non-oscillatory near shocks while maintaining second-order accuracy away from shocks. However, as we have already noted,  the quantity $|\partial_x u^\epsilon|$ may become highly irregular near shock discontinuities, and may thus
require setting the constant  $\beta \gg 1$ in order to  stabilize incipient numerical oscillations (see Section \ref{sec:sodshock} for evidence to this observation). While this increase in $\beta$ does not effect the asymptotic accuracy of the scheme, it is clearly beneficial to take $\beta$ as small as possible to preserve the correct amplitude and wave speed.

The loss of regularity of the coefficient  $|\partial_x u^\epsilon|$  suggests that a smoothed version of $|\partial_x u^\epsilon|$ would greatly benefit the dynamics.  Smoothing
$|\partial_x u^\epsilon|$ in space is not sufficient, as we must ensure smoothness in time as well.  Hence,
 we propose our $C$-method, which indeed provides a regularized version of \eqref{eqn:classicalViscosityEuler} and
allows for the use of much smaller values of  $\beta$ (less localized artificial dissipation), higher accuracy,  and practical viability.

%

\subsection{$C$-method for compressible Euler}
\label{subsec:cMethodEuler}

Analogous to \eqref{eqn:classicalViscosityEuler}, we control the amount of viscosity in \eqref{eqn:strongVanishingViscosity1} by the use of a function $C(x,t)$  of space and time. 
This function $C(x,t)$ is a solution to a reaction-diffusion equation, coupling to the evolution of $\u^\epsilon$. The mechanism of diffusion, smoothing/spreading the sharp peaks localized around discontinuities in the velocity competes with the mechanism of reaction, pushing $C(x,t)$ to zero. Subsequently, our $C$-method yields a smooth, yet  sharp, distribution of artificial viscosity yielding regularized $\u^\epsilon$ similar to \eqref{subeq:strongVanishingViscosity} on which we can build high-resolution numerical schemes.

For fixed ${\bf u}_0$ we choose $\beta > 0$. Then, for each $\epsilon > 0$, we find 
$$
{\bf u}^\epsilon(x,t) = \left ( \begin{array}{c} \rho^\epsilon(x,t) \\ m^\epsilon(x,t) \\ E^\epsilon(x,t) \end{array} \right ) \ \  \text{ and } \ \ C^\epsilon(x,t)
$$
as solutions of the following  parabolic system of (viscous) conservation laws:
\begin{subequations}
\label{subeq:cmethodEuler}
\begin{alignat}{2}
\partial_t {\bf u}^\epsilon + \partial_x {\bf F}({\bf u}^\epsilon) & = \partial_x \left(\tilde\beta \epsilon^2  C^{\epsilon, \delta }   \partial_x {\bf u}^\epsilon \right), && \quad t > 0 , \label{eqn:cmethodEuler1} \\
\partial_t C^\epsilon - S({\bf u}^\epsilon)\partial^2_x  C^\epsilon +    \frac{S({\bf u}^\epsilon)}{\epsilon} C^ \epsilon   & =  S({\bf u}^\epsilon ){G(\partial_xu^\epsilon)} 
 \, , && \quad t > 0, \label{eqn:cmethodEuler2} \\
\left ( {\bf u}^\epsilon , C^\epsilon \right )  & = ( {\bf u}^\epsilon_0, G(\partial_x u^\epsilon_0) ),&& \quad t = 0, \label{eqn:cmethodEuler3}
\end{alignat}
\end{subequations}
where $C^{\epsilon, \delta } = C^ \epsilon + \delta $ for a {\it fixed} positive constant $ 0 < \delta < \Delta x$, and
$\tilde \beta = \beta  \frac{\underset{\mathcal{I}}{\max} | \partial_x u^\epsilon|}{\underset{\mathcal{I}}{\max}C^\epsilon}$.
The forcing to equation (\ref{subeq:cmethodEuler}b) is defined as
\begin{equation}
\label{eqn:defineG}
G(\partial_x u^\epsilon) =  \frac{ |\partial_x u^\epsilon| }{\underset{\mathcal{I}}{\max} \ |\partial_x u^\epsilon|}
 \end{equation}
$({\bf u}^\epsilon)$ is defined by \eqref{subeq:maxWaveSpeed}, and ${\bf u}^\epsilon_0$ denotes a regularization of the initial data which we discuss below.
We also note that the scaling factor in $\tilde \beta$, given by
$\frac{\underset{\mathcal{I}}{\max} \  | \partial_x u^\epsilon|}{\underset{\mathcal{I}}{\max} \  C^\epsilon}$, is included only to make comparisons
with the classical artificial viscosity approach, but is in no way necessary.

\subsection{Regularization of initial data for use with FEM-C}

Unlike numerical algorithms which advance cell-averaged quantities, the finite-element method relies upon polynomial interpolation of
nodal values, and requires solutions to be continuous across element boundaries in order for the interpolation to converge.  As such,
the use of discontinuous initial data produces Gibbs-type oscillations, at least on very short time intervals.   To avoid this spurious behavior,
it is advantageous to smooth discontinuous initial profiles.

More specifically, we provide a hyperbolic-tangent smoothing for initial data ${\bf u}^\epsilon_0$ for our FEM-C scheme.
Since pointwise evaluation is well-defined for smooth functions, the finite-element discretization scheme can interpolate the regularized data and generate 
appropriate initial states.

For an interval $[a,b]$, we denote the indicator function  
\begin{equation}
\label{eqn:indicatorFunction}
{\bf 1}_{[a,b]}(x) = \begin{cases} 1 , \quad x \in [a,b],  \\ 0 , \quad  x  \notin [a,b] \,, \end{cases}
\end{equation}
and consider initial conditions with components of the form
\[
\left ({\bf u}^ \epsilon _0(x)\right )^i = \sum_{j=1}^{L_i} {\bf 1}_{[a^i_j,b^i_j]} (x) f^i_j(x),
\]
where the $[a^i_j,b^i_j]$ are pairwise-disjoint (in $i$) and each $f^i_j$ are smooth.

We then define the regularized initial condition
\[
\left ({\bf u}^\epsilon_0(x)\right )^i= \sum_{j=1}^{L_i} {\bf 1}^\epsilon_{[a^i_j,b^i_j]} (x) f^i_j(x),
\]
where
\[
{\bf 1}^\epsilon_{I^i_j}(x) = \frac{1}{2} \left [ \tanh \left ( \frac{x+b^i_j}{\epsilon} \right ) - \tanh \left ( \frac{x+a^i_j}{\epsilon} \right ) \right ].
\]

This regularization procedure achieves approximations of exponential-order away from discontinuities; near discontinuities,  it  is a first-order approximation, when measured in the $L^1$-norm. 
Specifically, if $\left ( {\bf u}_0 \right )^i$ is smooth in $\omega \subset \mathcal{I}$, then 
the $L^1(\omega)$-norm of the error 
\begin{equation}
\label{eqn:exponentialAccuracy}
\| \left ( {\bf u}_0 \right )^i - \left ({\bf u}_0^{\epsilon} \right )^i \|_{L^1(\omega)} = \int_\omega \left | \left ( {\bf u}_0(x) \right )^i - \left ( {\bf u}^\epsilon_0(x) \right)^i \right | \ dx = O(\epsilon^p)
\end{equation}
for any positive integer $p$.
Alternatively, if ${\bf u}^i_0$ is discontinuous somewhere in $\Omega \subset \mathcal{I}$, the $L^1(\Omega)$-norm of the error
\begin{equation}
\label{eqn:firstOrderAccuracy}
\| \left ({\bf u}_0^i \right ) - \left ( {\bf u}^\epsilon_0 \right)^i \|_{L^1(\Omega)} = O(\epsilon).
\end{equation}

These observations assert that our regularization of the initial data allows for higher-order approximation of the initial data and is analogous to the averaging procedure required by Majda \& Osher  \cite{Majda1977671}.

\subsection{A compressive modification of the forcing $G$ in the $C$-equation}
\label{subsec:modificationsForG}

The function $G$ in \eqref{eqn:defineG} is chosen in such a manner so that $C^\epsilon$ is large where there are sharp transitions in the velocity field $u^\epsilon(x,t)$. 
In compressive regions (i.e., where $\partial_x u < 0$ or $\partial_x u^\epsilon < 0$), sharp transitions over lengths of $O(\epsilon)$ correspond to shocks and artificial viscosity is required so that ${\bf u}^\epsilon$ remains smooth. In expansive regions, corresponding to rarefactions,  artificial viscosity is not generally necessary. 

These observations motivate the following alternative forcing  function:
\begin{equation}
\label{eqn:defineGcomp}
G_{comp}(\partial_x u^\epsilon)  = \frac{|\partial_x u^\epsilon|}{\underset{\mathcal{I}}{\max} \ |\partial_x u^\epsilon |} {\bf 1}_{(-\infty,0)} \left ( \partial_x u^\epsilon \right )
\end{equation}
where the indicator function ${\bf 1}_{(-\infty,0)}$ introduces viscosity only in regions of compression.

The ability to use such a switch is heavily dependent on the use of a space-time smoothing. Since the velocity in many numerical schemes may become oscillatory near shocks, such a switch can become discontinuous between adjacent cells/elements. However, the space-time nature  of  the $C$-equation resolves this issue, providing a smooth artificial viscosity profile.

This modified function $ G_{comp}$  typically increases accuracy in Euler simulations, but can lead to a loss of stabilization.   For our FEM-C approach, where the stabilizing effects of artificial viscosity are necessary to dampen noise, the use of $G_{comp}$ is restricted to the problems of Sod and Osher-Shu, which contain only moderately strong shocks.

\subsection{Moving to the discrete level}
\label{subsec:move2Discrete}

The use of the C-equation yields smooth solutions ${\bf u}^\epsilon$ and thus we expect  that a variety of  higher-order discretization techniques, with sufficiently small $\Delta t$ and $\Delta x$,  could provide accurate, non-oscillatory approximations. In our implementation, artificial viscosity spreads discontinuities  over  regions of size $O(\beta \epsilon)$. Thus, given a particular initial condition, final time, discretization scheme, etc., we choose $\beta > 0$ such that the scaling $\epsilon = \Delta x$ produces non-oscillatory profiles. 

We also note that the initial condition for $C$, given in \eqref{eqn:cmethodEuler3} is chosen so to guarantee the coefficients of diffusion in \eqref{eqn:cmethodEuler1} are smooth up to $t = 0$. Moreover, choosing such initial conditions for $C$ allows one to recover the classical artificial viscosity as $\epsilon \to 0$.  As stated, these initial conditions may require a smaller time-step (by a factor of $10$) for the first few time-steps. In practice, taking $C \equiv 0$ is an effective simplification to eliminate the need for smaller initial time-steps. Alternatively, we can solve an elliptic PDE for $C$ at the initial time and similarly eliminate that concern.

\subsection{The $C$-method under a Galilean-transformation}
We begin our discussion for the case of constant entropy.
The Galilean invariance of the isentropic Euler equations results from  the advective nature of the PDE. Since we solve a modified equation (coupled with the additional $C$-equation) it is of interest to know to what extent Galilean invariance is preserved. For simplicity, we assume that
\[
p(x,t) = \rho(x,t)^2 \,.
\]
 (The choice $\gamma=2$ corresponds to the shallow water equations, but 
any other choice of $\gamma >1$ works in the same fashion.)

Given a fixed $v \in \mathbb{R}$ we define the change in independent variables
\[
\tilde x = x - vt , \quad \tilde t  = t,
\]
denoting $\phi(\tilde x,\tilde t) = (x,t)$
and the analogous change in the dependent variables 
\begin{equation}
\label{eqn:rhoAndu}
\tilde \rho(\tilde x, \tilde t) = \rho(\tilde x + v \tilde t,\tilde t), \quad \tilde u(\tilde x, \tilde t) = u(\tilde x + v \tilde t,\tilde t) - v.
\end{equation}
A simple calculation yields
\begin{subequations}
\label{eulergaltransform}
\begin{alignat}{2}
\partial_{\tilde t} \tilde \rho  + \partial_{\tilde x} (\tilde \rho \tilde u) & = \left [\partial_t \rho + \partial_x (\rho u) \right ] \circ \phi, && \\
\partial_{\tilde t} (\tilde \rho \tilde u) + \partial_{\tilde x} (\tilde \rho \tilde u^2 + \tilde p) & = \left [ \partial_t (\rho u) + \partial_x (\rho u^2) \right ] \circ \phi + \partial_{\tilde x} \tilde p - v \left [\partial_t \rho + \partial_x (\rho u) \right] \circ \phi.
\end{alignat}
\end{subequations}
We further have that
\begin{equation}
\label{eqn:pre}
\tilde p (\tilde x, \tilde t) = p(\tilde x + v \tilde t, \tilde t),
\end{equation}
so that the mass and momentum equations are, in fact, Galilean invariant in the absence of artificial viscosity.

With the $C$-method employed, \eqref{eulergaltransform} transforms to
\begin{subequations}
\label{eulergaltransformwithC}
\begin{alignat}{2}
\partial_{\tilde t} \tilde \rho  + \partial_{\tilde x} (\tilde \rho \tilde u) & = \left [\partial_x ( \mathcal{C} \partial_x \rho )\right ] \circ \phi, && \\
\partial_{\tilde t} (\tilde \rho \tilde u) + \partial_{\tilde x} (\tilde \rho \tilde u^2 + \tilde p) & = \left ( \partial_x \{ \mathcal{C} \partial_x [\rho (u-v)] \} \right ) \circ \phi ,
\end{alignat}
\end{subequations}
where we let $\mathcal{C} = \epsilon^2  \tilde \beta  C$,  and drop the $\epsilon$ superscript for notational convenience.

Examining \eqref{eqn:cmethodEuler2}, we see that the equation for $C$ is not Galilean invariant, but this is not a physical quantity, but
can rather be viewed as a parameter to the modified system of conservation laws.   As such we  {\it define} the behavior of $C$ under
Galilean transformations as follows:
\[
 \tilde{\mathcal{C}} (\tilde x, \tilde t) = \mathcal{C} (\tilde x + v \tilde t, \tilde t).
\]

With this definition of $\tilde C$, we find that
\begin{subequations}
\label{eulergaltransformwithC2}
\begin{alignat}{2}
\partial_{\tilde t} \tilde \rho  + \partial_{\tilde x} (\tilde \rho \tilde u) & = \left [\partial_{\tilde x} ( \tilde{\mathcal{C}} \partial_{\tilde x} \tilde \rho )\right ], && \\
\partial_{\tilde t} (\tilde \rho \tilde u) + \partial_{\tilde x} (\tilde \rho \tilde u^2 + \tilde p) & = \left \{ \partial_{\tilde x} [ \tilde{\mathcal{C}} \partial_{\tilde x} (\tilde \rho \tilde u) ] \right \} ,
\end{alignat}
\end{subequations}
and hence the $C$-method for isentropic Euler retains the  Galilean invariance.

We remark that in the absence of artificial viscosity on the right-hand side of the mass equation, the artificial flux term in the momentum equation is modified according to (\ref{ss3}) below.
This modification ensure Galilean invariance when the mass equation is left unchanged, which is the strategy employed for our WENO-C scheme.

Next, since the Galilean symmetry is for the smooth solutions (for which classical derivatives are well-defined), and since smooth velocity fields simply transport the
entropy function, it is thus a consequence of the transport of entropy, that Galilean invariance holds for the non-isentropic case as well.    The important of a numerical
approximation to capture the Galilean invariant solution is fundamental to the initiation of the Kelvin-Helmholtz instability and other basic instabilities present
in the Euler equation; see Robertson, Kravtsov, Gnedin, and Rudd  \cite{RoKrGnAbRu2010} for a thorough discussion.  In this connection, we next examine long wavelength
instabilities which can arise for very slowly moving shock waves.

\subsection{Regularization through the C-equation}

\begin{figure}[htbp]
\subfigure[The $C$-equation with $G(u_x)$]{
\label{subfig:CGvsu_x}
\includegraphics[scale=0.27]{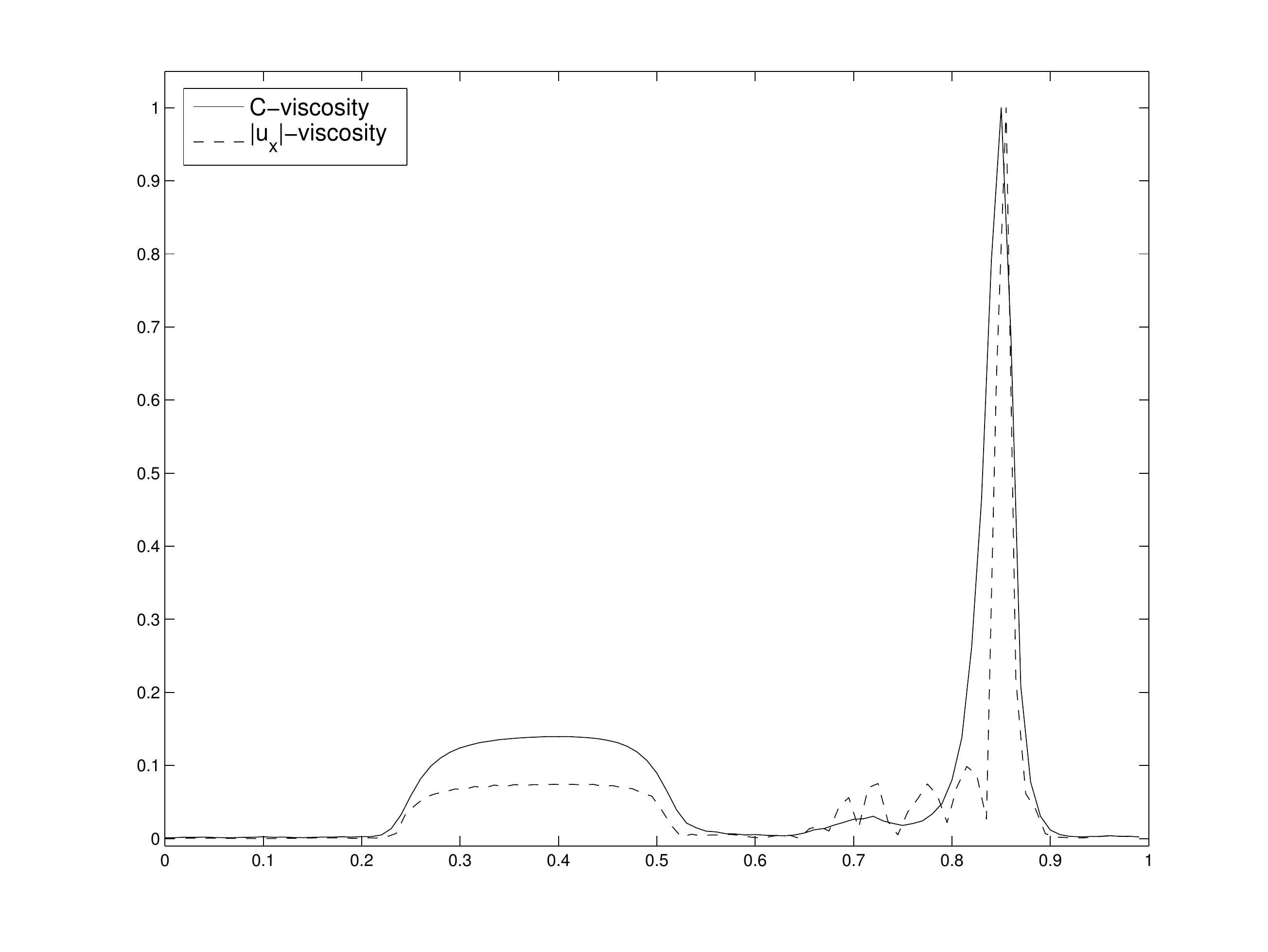}
}
\subfigure[The $C$-equation with $G_{comp}(u_x)$]{
\label{subfig:CGcompvsGcomp}
\includegraphics[scale=0.27]{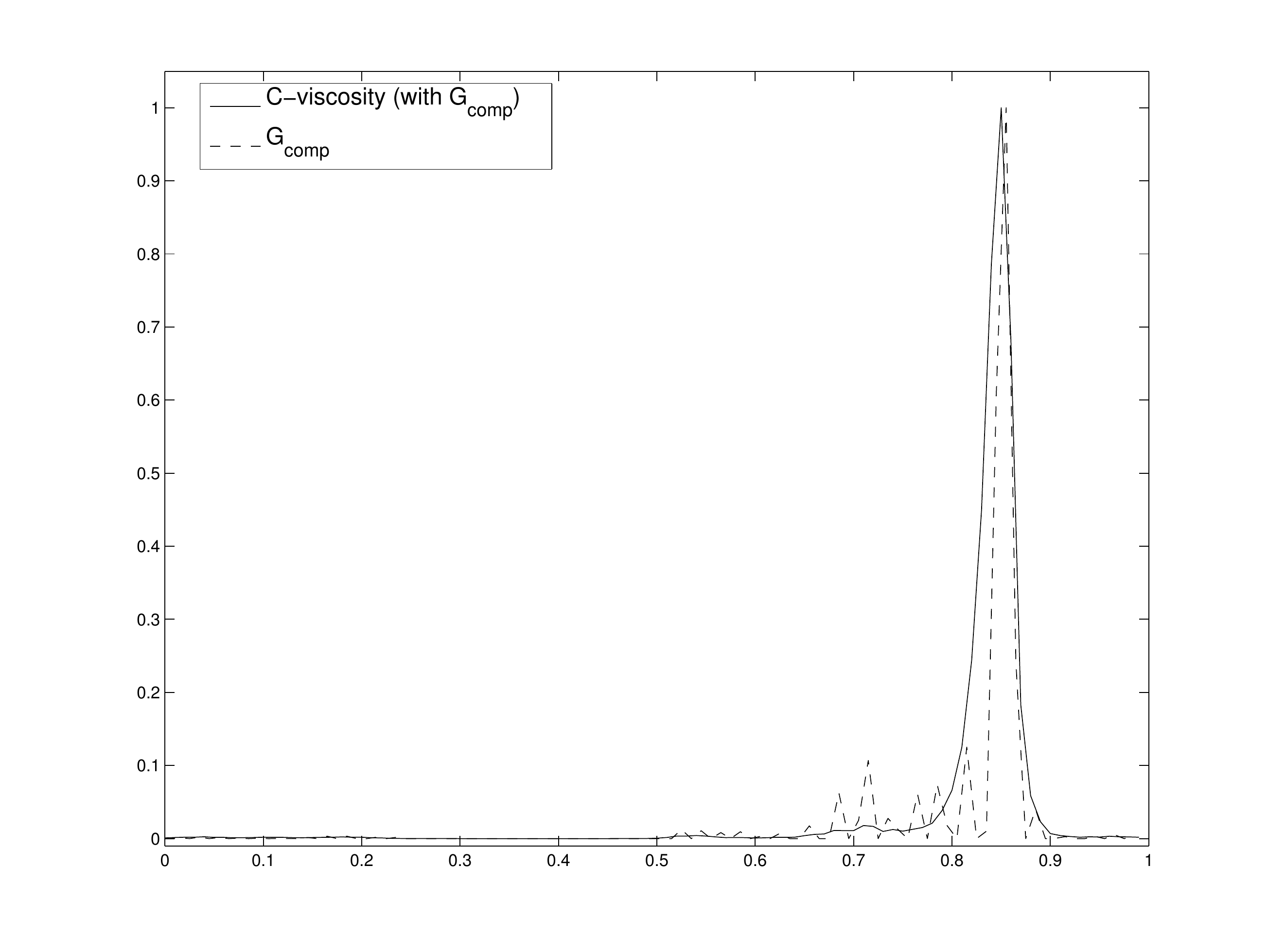}
}
\caption{A comparison of the artificial viscosity profile produced by the $C$-method and the classical Richtmyer-type approach for the Sod shock tube at $t=.2$.  Figures (a) and (b)
are with the compression switch off and on, respectively.  The smooth solid line is the $C$-method solution, while the oscillatory dashed line is the $|u_x|$-Richtmyer-type viscosity.}
\label{fig:spacetimesmoothing}
\end{figure}

It is of interest to examine the relative smoothness of $C$ to its unsmoothed counterpart $|u_x|$, and to determine the effect of this smoothing relative to the classical artificial viscosity approach. In Figure \ref{fig:spacetimesmoothing} we provide two plots demonstrating the effect of the $C$-method. In Figure \ref{subfig:CGvsu_x} we see that the $C$-equation provides a smoothened viscosity profile compared to the classical approach. Alternatively, in Figure \ref{subfig:CGcompvsGcomp} we plot $C$ using the compression-switch modification $G_{comp}$ versus using purely the quantity $G_{comp}$ (not smoothed by the $C$-equation) as a viscosity. In both cases we see how the $C$-method provides a far smoother  profile with roughly the same magnitude as the  non-smoothened approach. 

A useful feature of the $C$-method is the ability to tune parameters in the $C$-equation to generate non-oscillatory behavior. Though we are quite explicit on the form of the $C$-equation in  \eqref{eqn:cmethodEuler2}, a simple modification allows for the diffusion coefficient to be problem dependent, i.e. allowing for a choice of positive constant $\gamma > 0$ and replacing the diffusion term with
\[
- \gamma S({\bf u}^\epsilon)\partial^2_x C^\epsilon \,.
\]
In most of the forthcoming experiments, we fix $\gamma = 1$, but we note that choosing larger $\gamma$ can yield smoother solution profiles as the profile of $C$ will be less localized. 
The parameter $\gamma$ is a time-relaxation parameter, and can be viewed in an analogous fashion to the time-relaxation parameter present in Cahn-Hilliard and Ginzburg-Landau theories.
For very slow moving shocks, the time-relaxation can be adjusted to scale with the shock speed.\footnote{We note that $\gamma$ is  inversely proportional to the Mach number and its precise functional relation shall be examined in future work.}

We find this to be an effective approach for the flattening procedure discussed in \cite{Colella1984174} for removing  oscillations that form to the left of a slowly right-moving shock.  Moreover, Roberts
\cite{Roberts1990} concludes that a differentiable form of the numerical flux construction appears necessary to remove downstream long-wavelength oscillations caused by slow shock motion.  We use
the $C$-method to analyze this.

\begin{figure}[htbp]
\subfigure[$\gamma = 1$]{
\label{subfig:gam1}
\includegraphics[scale=0.27]{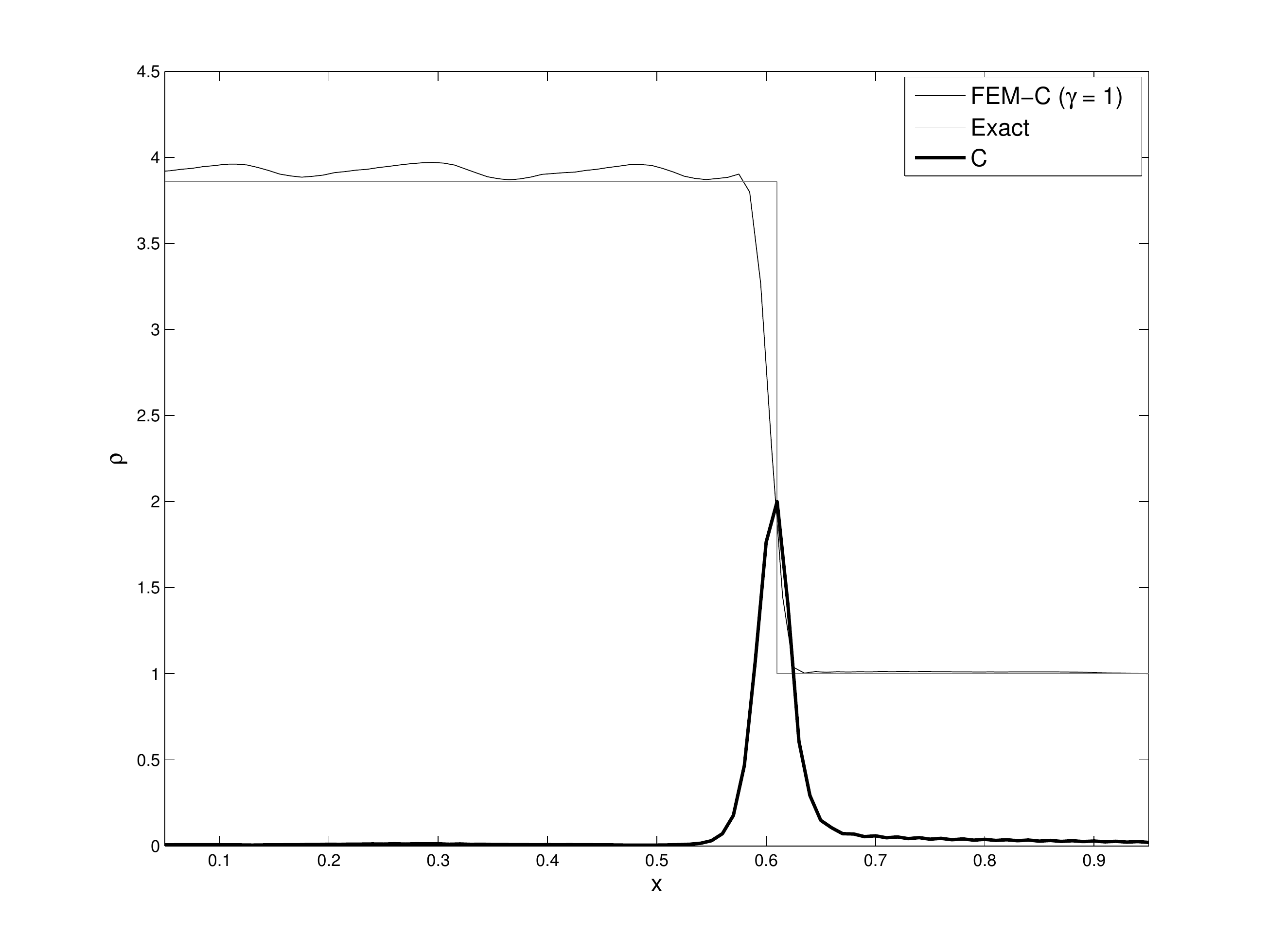}
}
\subfigure[$\gamma = 100$]{
\label{subfig:gam100}
\includegraphics[scale=0.27]{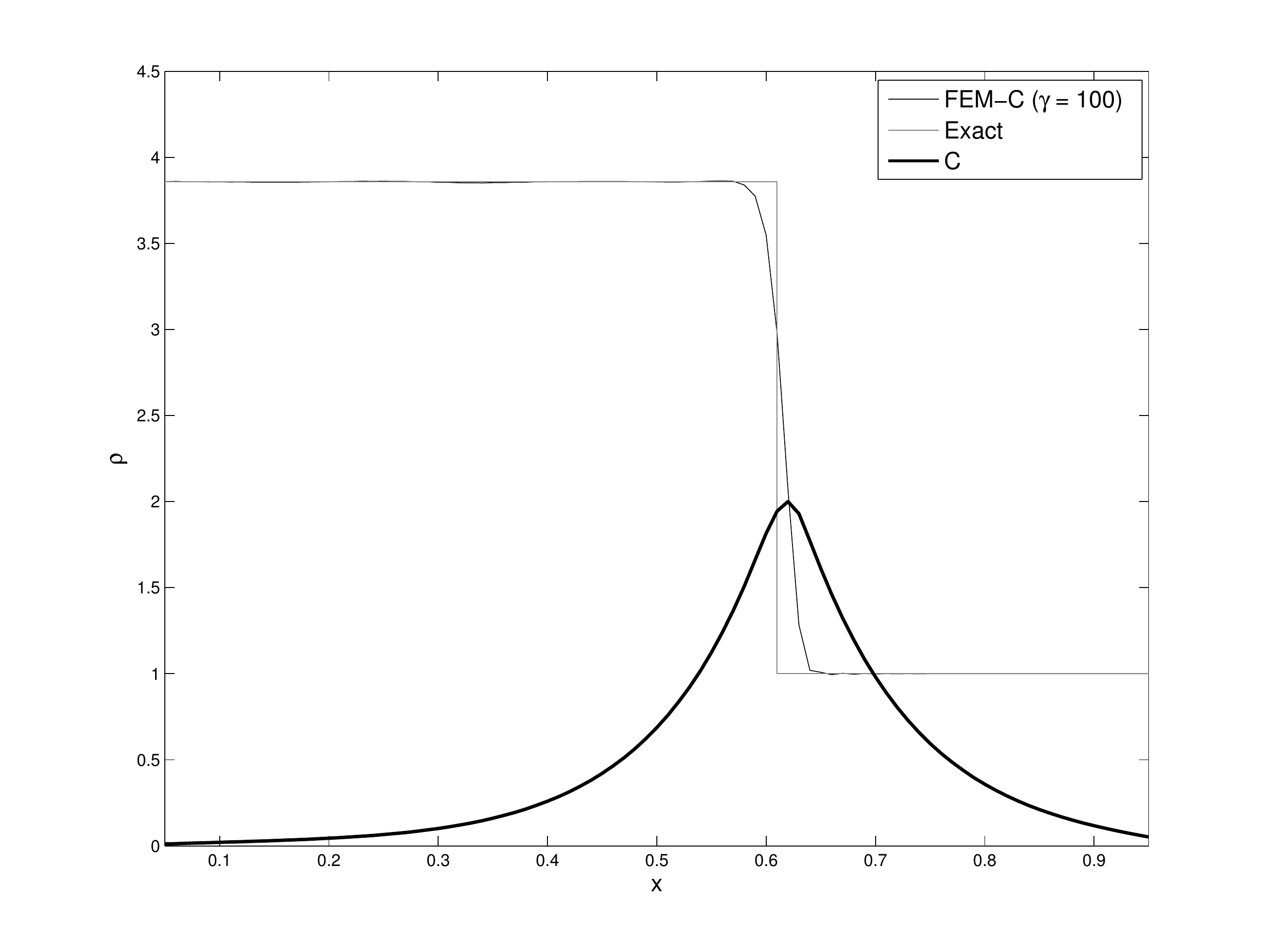}
}
\caption{Application of FEM-C to a very slowly moving shock}
\label{fig:slowshock}
\end{figure}

 Using the slow-shock initial conditions outlined in Quirk \cite{Quirk1994555}, in Figure \ref{fig:slowshock} we show the success of the FEM-C (outlined below in Section \ref{subsec:FEMC}) in removing these oscillations when choosing $\gamma = 1$ (Fig. \ref{subfig:gam1}) and $\gamma = 100$ (Fig. \ref{subfig:gam100}).   

\subsection{Convergence of the $C$-method in the limit of zero mesh size}
\subsubsection{The isentropic case}
We sketch the proof for the isentropic Euler equations given by
 \begin{subequations}
\label{isen}
\begin{alignat}{2}
(\rho u)_t+ (\rho u^2 +p)_x& = 0, &&  \\
\rho_t + (\rho u)_x &= 0 \,, && \\
p(\rho) & = \rho^  \gamma \,, &&
\end{alignat}
\end{subequations}
where $ \gamma >1$.  

To simplify the notation, we set $\tilde\beta=1$, and set the momentum $m^ \epsilon = \rho^ \epsilon u^ \epsilon $.  Following (\ref{subeq:cmethodEuler}), we write the $C$-method version of (\ref{isen}) as
 \begin{subequations}
\label{isenC}
\begin{alignat}{2}
m^ \epsilon _t+   [(m^ \epsilon )^2/ \rho^ \epsilon  +p^ \epsilon ]_x& =  \epsilon ^2 [C^ { \epsilon , \delta }  m^ \epsilon  _x]_x, &&  \\
\rho^ \epsilon _t + m^ \epsilon _x &=  \epsilon ^2 (C^ { \epsilon , \delta }  \rho_x)_x \,, && \\
p^ \epsilon (\rho^ \epsilon ) & =( \rho^ \epsilon ) ^  \gamma \,, && \\
C^ \epsilon _t -   S (u^ \epsilon )  C^ \epsilon _{xx} + \frac{S (u^ \epsilon )}{ \epsilon } C^ \epsilon  &= S(u^ \epsilon ) G(u^ \epsilon _x)\,, &&
\end{alignat}
\end{subequations}
or (\ref{isenC}a,b) can equivalently in terms of the vector ${\bf u^ \epsilon } = (m^\epsilon , \rho^\epsilon )$ and flux ${\bf f(u^ \epsilon )} = ( (m^ \epsilon )^2/ \rho^ \epsilon  +
(\rho^ \epsilon)^ \gamma  \ , \ m^ \epsilon )$ as
\begin{equation} 
{\bf u^ \epsilon }_t  + {\bf f(u^ \epsilon )}_x = \epsilon ^2 \left[ \mathcal{C} {\bf u^ \epsilon }_x\right]_x \,, \tag{\ref{isenC}'}
\end{equation} 
where $ \mathcal{C} $ denotes a diagonal 2x2 matrix with entries $C^{ \epsilon , \delta }$ which is strictly positive-definite.
Recall that
$G(u^ \epsilon )= |u^ \epsilon _x|/\max |u^ \epsilon _x|$, satisfies $G \ge 0$, and that $S(u^ \epsilon ) = \max (|u^ \epsilon +c|, |u^ \epsilon -c|)$, with $c$ denoting the sound speed.  On any time interval
$[0,T]$, the maximum wave speed $S(u^ \epsilon )$ is uniformly strictly positive; thus, as the initial data for $C^ \epsilon _{t=0} \ge 0$, the maximum principle shows that
$C^ \epsilon (x,t)$ must be non-negative.    We remark that while the use of $ C^{ \epsilon , \delta } = C^ \epsilon  + \delta $ as the coefficient is not required for the numerics, 
as  $ \delta $ is taken much smaller than the mesh size $\Delta x$,  strict positivity of 
$ \mathcal{C} $  simplifies the proof of regularity
of solutions to (\ref{isenC}) as well as the convergence argument.

 To avoid issues with spatial boundaries,  we shall assume periodic boundary conditions for our spatial domain.
Note that in this case, the fundamental theorem of calculus shows that $\frac{d}{dt} \int \rho (x,t) \,dx =0$ and that mass is conserved.

\subsubsection{The basic energy law}
In order to prove  that solutions to  (\ref{isenC}) converge to  solutions of   (\ref{isen}), we must  establish $ \epsilon $-independent estimates for solutions
of (\ref{isenC}).   To do so, we multiply equation (\ref{isenC}a) by $u^ \epsilon $,  integrate over our spatial domain, and make use of the
equation (\ref{isenC}b) to find that 
any  weak solution  to (\ref{isenC}) must verify the basic energy law
\begin{equation}\label{energylaw}
  \frac{d}{dt} \left[ \int  {\frac{1}{2}} \rho^ \epsilon  (u^ \epsilon )^2\, dx + {\frac{1}{\gamma -1}} \int p^ \epsilon \, dx\right] \le -  \epsilon ^2 \int  C^{ \epsilon , \delta } \rho^ \epsilon  \,  (u^ \epsilon _x)^2 \, dx -  \epsilon^2  \gamma \int C^{ \epsilon , \delta } (\rho^ \epsilon )^{\gamma-2} (\rho^ \epsilon _x)^2\, dx \,.
\end{equation} 
(The inequality in (\ref{energylaw}) is due to the lower semi-continuity of weak convergence and is replaced with equality for solutions which are sufficiently regular.)
Thus, the total energy of isentropic gas dynamics is dissipated according to the right-hand side of (\ref{energylaw}), and for each $ \epsilon >0$, we see that
$u^ \epsilon _x$ and $ \rho^ \epsilon _x$ are square-integrable (in $L^2$) for almost every instant of time, if the density $\rho^ \epsilon \ge \lambda >0$, that is, if $\rho^ \epsilon $
avoids vacuum.   We shall explain below that this is indeed the case.

\subsubsection{Regularity of solutions ${\bf u^ \epsilon }$}  The reaction-diffusion equation (\ref{isenC}d) is a uniformly parabolic equation.   According to the
energy law, and as a consequence of Sobolev's theorem, $u^ \epsilon $ is a bounded function; furthermore,  the right-hand side of (\ref{isenC}d) is in $L^2$.
It is standard, from the regularity theory of uniformly parabolic equations, that $C^ \epsilon $ then has two spatial (weak) derivatives which are square-integrable.   This,
in turn, shows that for $ \epsilon >0$, solutions ${\bf u^ \epsilon }$ possess three spatial (weak) derivative which are square-integrable for almost every instant of time.  This
implies that solutions ${\bf u^ \epsilon}$ are classically differentiable in both space and time.

Furthermore, by using the symmetrizing matrix 
$\left[
\begin{array}{lr}
\rho^ \epsilon  & 0 \\
0& \gamma (\rho^ \epsilon )^{\gamma -2})
\end{array}\right]
$
we can show that $(u^ \epsilon( \cdot , t) , \rho^ \epsilon(\cdot ,t) )$ are, independently of $ \epsilon $ and $t$,  uniformly bounded in the Sobolev space $H^2$ (consisting of measurable functions
with two weak derivatives in $L^2$), and thus we may take a pointwise limit of this sequence as $ \epsilon \to 0$,  in the event that the time-interval is sufficiently small
as to ensure that a shock has not yet formed.   Of course, we are interested, in convergence to discontinuous profiles, so we address this next.

\subsubsection{Convergence to the entropy solution} \label{sec::entropy} We shall now provide a sketch of the limit as $ \epsilon  \to 0$.  A function $\eta: \mathbb{R}  ^2 \to \mathbb{R}  $ is called
an {\it entropy} for (\ref{isen}) with {\it entropy flux} $q: \mathbb{R}  ^2 \to \mathbb{R}  $ if smooth solutions ${\bf u}$ satisfy the additional conservation law
\begin{equation}\label{entropy}
\eta({\bf u})_t + q({\bf u})_x =0 \,.
\end{equation} 
In non-conservative form, (\ref{isen}) and (\ref{entropy}) are written as
$$
{\bf u} _t + \nabla f( {\bf u} ) {\bf u} _x =0 \,, \ \ \nabla \eta( {\bf u} ) {\bf u} _t + \nabla q ( {\bf u} ) {\bf u} _x =0 \,,
$$
from which we obtain the compatibility condition between $\eta$ and $q$,
\begin{equation}\label{comp_cond}
\nabla \eta( {\bf u} ) \, \nabla f ( {\bf u} ) = \nabla q ( {\bf u} ) \,.
\end{equation} 
The pair $(\eta, q)$ satisfy (\ref{entropy}) if and only if condition (\ref{comp_cond}) holds.    Moreover, a weak solution to (\ref{isen}) is the unique entropy solution if
\begin{equation}\label{entropy_ineq}
\eta({\bf u})_t + q({\bf u})_x  \le 0 \,.
\end{equation} 
For isentropic gas dynamics we can set
$$
\eta(m, \rho) = \frac{m^2}{2\rho} + \frac{\rho^ \gamma }{\gamma-1} 
$$
which is the total energy, with corresponding entropy flux 
$$
q(m, \rho) = \left[ \frac{m^2}{2\rho} + \frac{\gamma}{\gamma-1} \rho^ \gamma  \right] \frac{m}{\rho} \,.
$$
We observe that $ \nabla ^2 \eta (m, \rho)$ is strongly convex as long as $\rho > 0$.

For the sequence of solution $ {\bf u^ \epsilon }$ of (\ref{isenC}), suppose that as $ \epsilon \to 0$, $ {\bf u} ^\epsilon $ converges boundedly (almost everywhere) to 
a weak solution $ {\bf u} $ of (\ref{isen}).  We claim that if $(\eta, q)$ satisfy (\ref{entropy}), then  (\ref{entropy_ineq}) holds in the distributional sense.   To see that this
is the case, we take the inner-product of
$ \nabla \eta ( {\bf u} ^ \epsilon )$ with equation (\ref{isenC}'), and find that
\begin{align*} 
\eta({\bf u}^ \epsilon )_t + q({\bf u}^ \epsilon )_x &  = \epsilon ^2  \nabla \eta ( {\bf u} ^ \epsilon ) \left[ \mathcal{C} {\bf u}^ \epsilon _x\right]_x  \\
& =   \epsilon ^2 \left[ \mathcal{C} \eta({\bf u}^ \epsilon )_x\right]_x - \epsilon ^2  [{\bf u}^ \epsilon _x ]^T \, \mathcal{C} \, \nabla ^2 \eta
 ( {\bf u} ^ \epsilon ) {\bf u}^ \epsilon _x \label{etaeq} \,.
\end{align*} 
Integrating  over the spatial domain and then
over the time interval $[0,T]$ yields
$$
\int \eta ( {\bf u} ^ \epsilon (x,T)) dx - \int \eta ( {\bf u} ^ \epsilon (x,0)) dx = - \epsilon ^2 \int_0^T \int  [{\bf u}^ \epsilon _x ]^T \, \mathcal{C} \, \nabla ^2 \eta
 ( {\bf u} ^ \epsilon ) {\bf u}^ \epsilon _x  \, dx \, dt \,,
$$
from which it follows that
\begin{equation}\label{ss1} 
 \int_0^T \int  | \epsilon  {\bf u}^ \epsilon _x|^2  \, dx \, dt \le \bar c
\end{equation} 
where the constant  $\bar c$ depends upon $ \delta$, the minimum value of density, and the entropy in the initial data.   For 
a smooth, non-negative test function $ \psi$ with compact support in the strip $ \mathcal{I}\times (0,T)$,
$$
\iint \eta( {\bf u}^ \epsilon \phi_t + q( {\bf u} ^ \epsilon ) \phi_x \, dx dt = \epsilon\iint \mathcal{C} ( \epsilon {\bf u} ^ \epsilon )_x \phi_x \, dx dt 
+ \iint \epsilon ^2  [{\bf u}^ \epsilon _x ]^T \, \mathcal{C} \, \nabla ^2 \eta
 ( {\bf u} ^ \epsilon ) {\bf u}^ \epsilon _x \phi \, dx dt \,.
$$
Thanks to (\ref{ss1}), the first term on the right-hand side goes to zero like $ \epsilon $, while the second term is non-negative, since
$ \nabla ^2 \eta ( {\bf u} ^\epsilon )$ is positive-definite (since $\eta$ is strongly convex) as is $ \mathcal{C} $.  Thus, as $ \epsilon \to 0$,
we recover the entropy inequality (\ref{entropy_ineq}).

It remains to discuss the assumptions concerning the bounded convergence of $ {\bf u} ^ \epsilon $ to $ {\bf u} $, as well as the uniform bound from below
 $\rho^ \epsilon \ge \nu > 0$.      The argument relies on finding a priori bounds on the amplitudes of solutions to (\ref{isenC}).  If it is the case that
 uniformly in $ \epsilon >0$, 
 $$
 | {\bf u} ^ \epsilon | \le M  \text{  and  }  0< \nu \le  \rho^ \epsilon \,,
 $$ 
 then the compensated-compactness approach for isentropic Euler pioneered by DiPerna \cite{DiPerna1983}  and  made much more general by 
 Lions, Perthame, \& Souganidis \cite{LiPeSo1996} provides a subsequence of $ {\bf u} ^ \epsilon $ converging pointwise (almost everywhere) to
 a solution $ {\bf u} $ of (\ref{isen}).

For isentropic gas dynamics, our approximation (\ref{isenC}) preserves the invariant quadrants of the inviscid dynamics (just as in the case of uniform
artificial viscosity) and provides the bound $| {\bf u} ^ \epsilon | \le M$ as long as $ 0< \nu \le  \rho^ \epsilon $ for some $\nu$.  In particular, the Riemann
invariants $w = u + \frac{2\gamma}{\gamma-1} \rho^{ \sqrt{\gamma-1}}$ and $z = u - \frac{2\gamma}{\gamma-1} \rho^{ \sqrt{\gamma-1}}$ satisfy
$w(x,t) \le \sup w|_{t=0}$ and $-z(x,t) \le \sup (-z_{t=0})$ and the intersection of these half-planes provides the invariant quadrant (see Chueh, Conley, \& Smoller \cite{ChCoSm1977}), and hence
the desired bound $| {\bf u} ^ \epsilon | \le M$ as long as vacuum is avoided.

Finally, the fact that we have the lower-bound $ 0< \nu \le  \rho^ \epsilon $ is an immediate consequence of the strong maximum principle.

\subsection{The $C$-equation as a gradient flow}
Notice that equilibrium solutions to the $C$-equation are minimizers of the following functional (we drop the superscript $ \epsilon $):
\begin{equation}\nonumber
\mathcal{E}_G( C) = \int  \left( {\frac{1}{2}} C_x^2  - G(u_x) C + {\frac{1}{2 \epsilon }}  C^2\right) \, dx \,.
\end{equation} 

In the absence of a forcing function $G(u_x)$, this reduces to  
\begin{equation}\label{ss7}
\mathcal{E}_0( C) = {\frac{1}{2}}  \int  \left({C_x^2}   + {\frac{1}{ \epsilon }}  C^2\right) \, dx \,.
\end{equation} 
The first term is commonly referred to as the Dirichlet energy and its minimizers are harmonic functions.  The second term can be viewed as 
a {\it penalization of the Dirichlet energy}.    In particular,  because the {\it energy} functional is bounded by a constant independent of $ \epsilon >0$,
the penalization term constrains $C$ to be $O( \sqrt{ \epsilon })$.   Thus, minimizers are trying to be harmonic while minimizing their support.

The $C$-equation can be written as a classical gradient flow equation
$$
\frac{ dC}{dt} = - S(u) \nabla \mathcal{E} _G(C) \,,
$$
where the gradient is computed relative to the $L^2$ inner-product.    Thus the heat operator in the $C$-equation, $\partial_t - \partial^2_x$, smooths the forcing in space-time,
while the reaction term $S(u)/ \epsilon C$ minimizes the support of the smoothed profile.   This is very much related to the theories of Cahn-Hilliard and Ginzburg-Landau
gradient flows, and we intend to examine this connection in subsequent papers.

\section{Numerical Schemes}
\label{sec:numSchemes}

We describe two very different numerical algorithms in the context of our  $C$-method. First, we outline a classical continuous finite-element discretization, yielding FEM-C and FEM-$|u_x|$ (based on classical artificial viscosity). Second, we discuss a simple WENO-based scheme for compressible Euler that upwinds solely based on the sign of the velocity $u$. To this scheme, we apply a slightly modified $C$-method resulting in our WENO-C algorithm.

For the purpose of comparison, we also implement two additional numerical methods.
The first is a second-order central-differencing scheme of Nessayhu-Tadmor (NT), a nice and  simple method which serves as a base-line for our FEM-C comparisons. The second scheme is a very competitive WENO scheme that utilizes a Godunov-based upwinding based upon characteristic decompositions (WENO-G).  This will serve as a benchmark for our WENO-C scheme.

\subsection{Notation for discrete solutions} 

To compute approximations to \eqref{subeq:consLaw}, we subdivide space-time into a collection of spatial nodes $\{x_i\}$ and temporal nodes $\{t_n\}$. We denote the computed approximate solution by 
$$ {\bf u}^n_i \approx {\bf u}(x_i,t_n),$$
noting that for fixed $i$ and $n$, ${\bf u}^n_i$ is a 3-vector of solution components, i.e.,
\[
{\bf u}^n_i = \left [ \begin{array}{c} \rho^n_i \\ m^n_i \\ E^n_i \end{array} \right ].
\]

It is important to note that we
use the notation ${\bf u}^n_i$ for both pointwise approximations to ${\bf u}$, (acquired via  FEM-C) and approximations to the cell-average values of ${\bf u}$ (acquired via WENO-C).

A subscripted quantity $w_i$ denotes the vector itself {\em and} the individual components of the vector. We overload this notation so to not cause any confusion between functions defined over a continuum versus those defined only at a finite number of points.

In FEM-C and WENO-C, we discretize \eqref{subeq:cmethodEuler} (or some slight modification) with $\epsilon = \Delta x$, and use the above notation for the computed solution. We also denote the approximation to $C$ by $C^n_i$.

\subsection{FEM-C and FEM-$|u_x|$: A Second-Order Continuous-Galerkin Finite-Element Scheme}
\label{subsec:FEMC}

We choose a  second-order continuous-Galerkin finite-element scheme to provide a discretization  of \eqref{subeq:cmethodEuler}, subsequently defining our FEM-C scheme. 

We subdivide $\mathcal{I}$ with $N+1$ (for $N$ even)-uniformly spaced nodes $\{x_i\}$ separated by a distance $\Delta x$. In the FEM community, spatial discretization size is more commonly referred by {\em element-width}; to maintain consistency with the literature, we refer to the inter-nodal regions as {\em cells}. Since we use a continuous FEM, the degrees-of-freedom are defined at the cell-edges (as opposed to cell-centers)\footnote{When we compare  our FEM-C scheme with other, {\em cell-averaged} schemes, we perform an averaging procedure based upon averages between nodes.}.

For use in our FEM implementation, it is useful to consider the variational form of \eqref{subeq:cmethodEuler}. At the continuum level, $({\bf u}^\epsilon,C^\epsilon)$  satisfy
\begin{subequations}
\begin{equation}
\label{eqn:var1}
\int_\mathcal{I}\left[ \partial_t {\bf u}^\epsilon \cdot \Phi - {\bf F}({\bf u}^\epsilon) \cdot \partial_x \Phi + \beta \epsilon^2 {\frac{\underset{\mathcal{I}}{\max} \  |\partial_x u^\epsilon|}{\underset{\mathcal{I}}{\max} \  C^\epsilon}} C^\epsilon \partial_x {\bf u}^\epsilon \cdot \partial_x \Phi\right] \, dx = 0\,,
\end{equation}
\begin{equation}
\label{eqn:var2}
\int_\mathcal{I}\left[ \partial_t C^\epsilon \phi + S({\bf u^\epsilon}) \left ( \epsilon \partial_x C^\epsilon \partial_x \phi  + \frac{1}{\epsilon} C^\epsilon \phi \right ) \right]\,dx = \int_\mathcal{I} S({\bf u}^\epsilon)G(\partial_x u^\epsilon) \phi \ dx 
\end{equation}
\end{subequations}
for almost every $t$, for all vector-valued test functions $\Phi$, and all scalar-valued test functions $\phi$.

Using the finite-element spatial discretization  based on piecewise second-order Lagrange polynomials, we construct operators $\mathcal{A}_{\text{FEM}}$ and $\mathcal{B}_{\text{FEM}}$, corresponding to  the non-time-differentiated terms in \eqref{eqn:var1} and \eqref{eqn:var2}, respectively. 
Using these discrete operators, we write the semi-discrete form of (\ref{eqn:var1}) and (\ref{eqn:var2}) as 
\begin{equation}
\label{eqn:FEMCSemiDiscrete}
\partial_t \left [ \begin{array}{c} {\bf u}_i \\ C_i \end{array} \right ] + \left [\begin{array}{c} \mathcal{A}_{\text{FEM}} ({\bf u}_i, C_i) \\ \mathcal{B}_{\text{FEM}} ({\bf u}_i, C_i) \end{array} \right ] = 0
\end{equation}
where ${\bf u}_i$ and $C_i$ represent the nodal values of an approximation to ${\bf u}^\epsilon$ and $C^\epsilon$ for which $\epsilon = \Delta x$ (see Section \ref{subsec:move2Discrete}). For a standard reference on the details of this procedure, see Larsson \& Thom{\'e}e \cite{Larsson2003}.

The time-differentiation in \eqref{eqn:FEMCSemiDiscrete} is approximated by a diagonally-implict second-order time-stepping procedure; first we predict ${\bf u}_i^{n+1}$ to and solve the implicit set of equations for $C_i^{n+1}$ and follow by implicitly solving for ${\bf u}_i^{n+1}$ using $C_i^{n+1}$. Our fully discrete scheme is given by
\begin{subequations}
\label{subeq:FEM}
\begin{alignat}{2}
{\bf \tilde u}_i^{n+1} &= {\bf u}_i^{n} +  \mathcal{A}_{FEM}({\bf u}_i^n,C_i^n) , && \\ 
C_i^{n+1}  &= C_i^n + \frac{t_{n+1}-t_n}{2} \left [ \mathcal{B}_{FEM}({\bf \tilde u}_i^{n+1},C_i^{n+1}) + \mathcal{B}_{FEM}({\bf u}_i^n,C_i^n) \right ], && \\
{\bf u}_i^{n+1} &= {\bf u}_i^n + \frac{t_{n+1}-t_n}{2} \left [ \mathcal{A}_{FEM}({\bf u}_i^{n+1},C_i^{n+1}) + \mathcal{A}_{FEM}({\bf u}_i^n,C_i^n) \right ]. && 
\end{alignat}
\end{subequations}

For smooth solutions, where artificial viscosity is not necessary, our FEM-C scheme is second-order accurate in both space and time when the error is measured in the $L^1$-norm. Moreover, the addition the artificial viscosity obtained through the $C$-method is formally a second-order perturbation (in $\Delta x$) and we have verified this accuracy  when $\beta > 0$ (again, for smooth ${\bf u}_0$). For ${\bf u}_0$ containing jump discontinuities, the given scheme is no longer second-order accurate on all of $\mathcal{I}$ but preserves second-order accuracy in the smooth regions away from discontinuities.

For the classical artificial viscosity schemes \eqref{eqn:classicalViscosityEuler}, the C-equation is no longer used but we require a similar step to predict the velocity for use in the diffusion coefficient. This analogous scheme, is referred to as the FEM-$|u_x|$ scheme.

\subsection{WENO-C: A Simple WENO scheme using the $C$-method}

Our WENO-based scheme is motivated by Leonard's finite volume schemes (\cite{Leonard}, pg. 65). Upwinding is performed solely based on the sign of the velocity at cell-edges, and the WENO reconstruction procedure is formally fifth-order.

We divide the interval $\mathcal{I}$ into $N$ equally sized cells of width $\Delta x$, identifying the $N$ degrees-of-freedom as cell-averages over cells centered at the  $x_i$. The cell edges are denoted using the fraction index, i.e. 
\[
x_{i+1/2} = \frac{x_i+x_{i+1}}{2}
\]
Subsequently, we denote a cell-averaged quantity by $w_i$ and its values at the left and right endpoints by $w_{i-1/2}$ and $w_{i+1/2}$, respectively. 

Given a vector $w_i$, corresponding to cell-average values,  and vectors $z_{i-1/2}$, $z_{i+1/2}$ corresponding to left and right cell-edge values, we define the $j$th component of vector 
\[
\left [ \text{WENO}(w_i,z_{i \pm 1/2}) \right]_j = \frac{1}{\Delta x} \left ( \tilde w_{j+1/2} z_{j+1/2} - \tilde w_{j-1/2} z_{j-1/2} \right) 
\]
where the cell-edge values of $\tilde w_{j+1/2}$ are calculated using a fifth-order WENO reconstruction, upwinding based upon the sign of $z_{j+1/2}$.

For the flux in the energy equation, we use
\begin{multline}
\left [ \text{WENO}_E(E_i,u_{i \pm 1/2}) \right]_j = \frac{1}{\Delta x} \Big ( \tilde E_{j+1/2} u_{j+1/2} \frac{(1+\frac{p_j}{E_j}) + (1 +\frac{p_{j+1}}{E_{j+1}}) }{2} - \\\tilde E_{j-1/2} u_{j-1/2} \frac{(1+\frac{p_{j-1}}{E_{j-1}}) + (1 +\frac{p_{j}}{E_{j}}) }{2}  \Big ).
\end{multline}

Using this simplified WENO-based reconstruction, we construct the operators $\mathcal{A}_{\text{WENO}}$ and $\mathcal{B}_{\text{WENO}}$ where

\begin{subequations}
\label{subeq:ABWENO}
\begin{equation}
\label{eqn:AWENO}
\left [ \mathcal{A}_{\text{WENO}}  \left ( \left [ \begin{array}{c} \rho_i \\ m_i \\ E_i  \end{array} \right ], \ C_i \right ) \right ] = \left [ \begin{array}{c} \text{WENO} (\rho_i, u_{i \pm 1/2 } ) \\ 
\text{WENO}(m_i , u_{i \pm 1/2} ) + \tilde \partial p_i - {\frac{ \tilde \partial_C u_{i+1/2} - \tilde \partial_C u_{i-1/2}}{\Delta x} } \\
\text{WENO}_E(E_i, u_{i \pm 1/2}) 
\end{array} \right ] 
\end{equation}
\begin{equation}
\label{eqn:BWENO}
\mathcal{B}_{\text{WENO}} \left ( \left [ \begin{array}{c} \rho_i \\ m_i \\ E_i  \end{array} \right ], \ C_i \right )  = - \frac{S({\bf u}_i)}{\Delta x} \left [C_i - G(\tilde \partial u_i) \right ] + \frac{\tilde \partial_S C_{i+1/2} - \tilde \partial_S C_{i-1/2} }{\Delta x}.
\end{equation}
\end{subequations}
where for a general quantity $w_i$, defined at the cell-centers, we denote
\[
w_{i+1/2} = \frac{w_{i+1} + w_{i} }{2}, \quad \tilde \partial w_i := \frac{w_{i+1} - w_{i-1}}{2\Delta x}, \quad \tilde \partial w_{i+1/2} = \frac{w_{i+1}-w_{i}}{\Delta x}.
\]
We also use the shorthand notation
\[
\tilde \partial_C u_{i+1/2} = \beta \ \Delta x^2 \ \max_{i} \left | \tilde \partial u_{i+1/2} \right | \  \frac{C_{i+1/2}}{\underset{i}{\max}  \ C_i} \ \rho_{i+1/2} \  \tilde \partial u_{i+1/2},
\]
and 
\[
\tilde \partial_S C = \Delta x \ S({\bf u}_i) \ \tilde \partial C_{i+1/2}.
\]

Using the above definitions, we  define the semi-discrete form
\begin{equation}
\label{eqn:wenoCSemiDiscrete}
\partial_t \left [ \begin{array}{c} {\bf u}_i \\ C_i \end{array} \right ] + \frac{1}{\Delta x} \left [\begin{array}{c} \mathcal{A}_{\text{WENO}} ({\bf u}_i, C_i) \\ \mathcal{B}_{\text{WENO}} ({\bf u}_i, C_i) \end{array} \right ] = 0
\end{equation}
and we generate the sequence of iterates ${\bf u}^n_i$ and $C^n_i$ with a standard fourth-order Runge-Kutta time-stepper.

The  resulting discretization outlined above is a slight variation on that outlined in \eqref{subeq:cmethodEuler}. While the amount of artificial viscosity $C(x,t)$ is controlled by only the velocity, we only add artificial viscosity to the momentum equation. This change is based upon the fact that WENO already minimizes the production of numerical oscillations and the addition of artificial viscosity is primarily intended on stabilizing the solution near strong shocks, whereas standalone WENO may  lose stability.  Without dissipation on the right-hand side of the mass equation, it is necessary to modify the artificial viscosity
on the momentum equation as follows:
\begin{equation}\label{ss3}
 \epsilon ^2 \tilde \beta \partial_x (  C \partial_x (\rho u) ) \to   \epsilon ^2 \tilde \beta \partial_x (C\rho \partial_x u ) \,.
\end{equation} 
This modification allows the $C$-method to maintain a basic energy law (in fact, it is the energy law (\ref{energylaw}) with the last term on the right-hand side), and simultaneously
permits higher accuracy for our WENO-based scheme.
 
\subsection{NT: A Second-order  Central-Differencing scheme of Nessayhu-Tadmor}

The central-differencing scheme of Nessyahu and Tadmor is an extension of the first-order Lax-Fredrichs finite difference scheme in which linear, MUSCL-based reconstructions are used to yield a second-order accurate scheme. The resulting scheme is extremely easy to implement (a FORTRAN code for 2-D problems is given in the Appendix of \cite{Jiang98}) and does not require the use of Riemann solvers or characteristic directions for the purpose of upwinding. The NT scheme allows for various choices of limiters to enforce TVD or ENO but the UNO-limiter (see Harten \& Osher \cite{Harten1987279}) is the most successful for our range of experiments.

Though that NT is easy to implement and is easy generalized to multi-D (yielding the JT-scheme \cite{Jiang98}),  it merely serves as a base-line comparison for our FEM-C. Both FEM-C and NT are second-order, but FEM-C turns out to be far less diffusive by comparison. 

\subsection{WENO-G: WENO with Godunov-based upwinding} 

In \cite{rideretal} the authors study a fifth-order, WENO-based discretization, upwinding by virtue of a high-accuracy Godunov-scheme. 
Their scheme has the usual trait of  WENO, offering minimal diffusion near extrema, and has the added  stabilization and accuracy of  higher-order Godunov solvers. For a more in-depth description, see \cite{rideretal}.

\begin{figure}[htbp]
\subfigure[Complete density profile]{
\label{subfig:sodshockCompareSameBetaZoomOut}
\includegraphics[scale = 0.27]{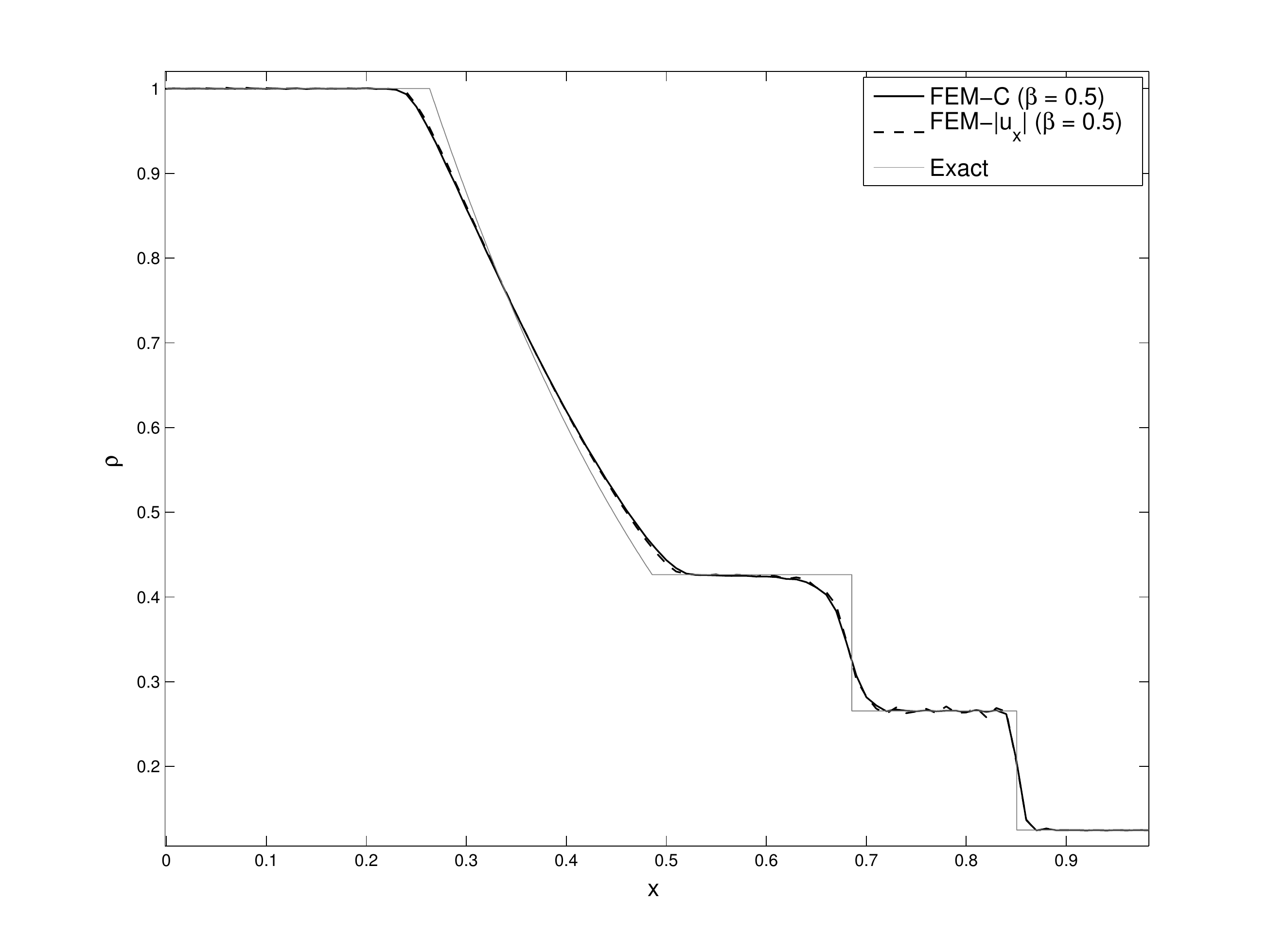}
}
\subfigure[Zoom-in at shock]{
\label{subfig:sodshockCompareSameBetaZoomIn}
\includegraphics[scale=0.27]{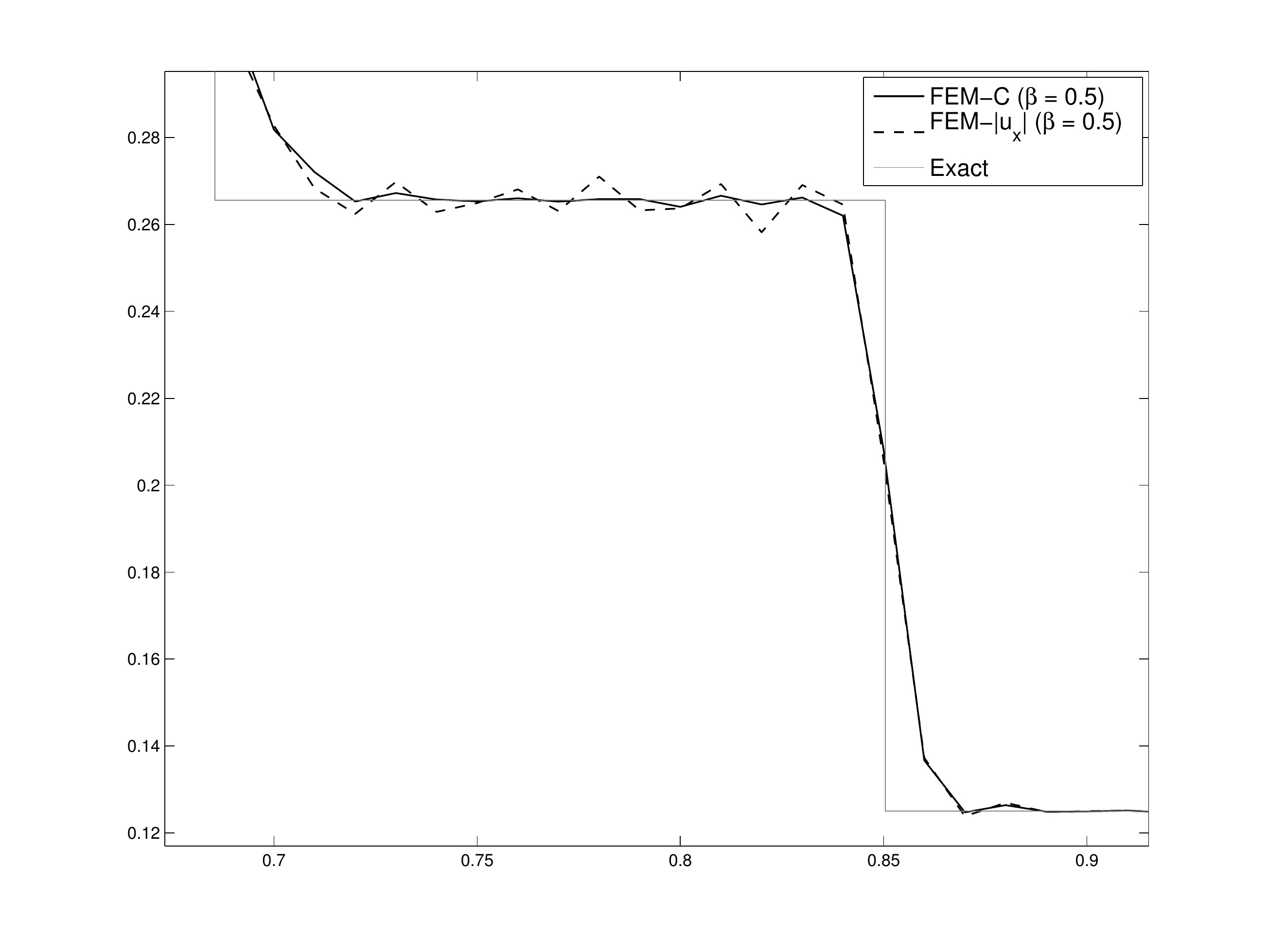}
}
\caption{Comparison of FEM-C and FEM-$|u_x|$, for the Sod shock-tube experiment with $N=100$, $T=0.2$. $\beta = 0.5$ for both FEM-C and FEM-$|u_x|$.}
\label{fig:sodshockCompareSameBeta}
\end{figure}

\section{Sod shock-tube problem}
\label{sec:sodshock}

For the classic Sod shock-tube problem, we consider the domain $\mathcal{I} = [0,1]$ along with the initial conditions
\begin{equation}
\label{eqn:sodshockIC}
\left ( \begin{array}{c} \rho_0(x) \\ m_0(x) \\ E_0(x) \end{array} \right ) = 
\left ( \begin{array}{c} 1 \\ 0 \\ 2.5 \end{array} \right ) {\bf 1}_{[0,\frac{1}{2})}(x) + 
\left ( \begin{array}{c} 0.125 \\ 0 \\ 0.25 \end{array} \right ) {\bf 1}_{ [\frac{1}{2},1]} (x),
\end{equation}
imposing natural boundary conditions at $x=0$ and $x=1$. This standard test problem, first considered in Sod \cite{Sod19781}, is a preliminary test for the viability of numerical schemes. An exact solution is known for this problem and consists of two nonlinear waves (one shock and one rarefaction) along with a contact discontinuity.

In Figure \ref{subfig:sodshockCompareSameBetaZoomIn} we compare the results of FEM-C and FEM-$|u_x|$ at $t=0.2$ using $N=100$ cells. We note that this comparison uses the standard choice of $G$ in \eqref{subeq:cmethodEuler} since we are merely concerned with the C-equation performing as a smooth version of classical artificial viscosity schemes. Unlike comparisons with the schemes based on cell-averages, we compare the {\em nodal} values of FEM-C and FEM-$|u_x|$. 
In this comparison, we choose $\beta = 0.5$ for both schemes
and see that the accuracy of both FEM-C and FEM-$|u_x|$ are quite comparable and each scheme resolves the shock in 3 cells. However, we notice noise in FEM-$|u_x|$ near the shock. In  Figure \ref{subfig:sodshockCompareSameBetaZoomIn} this observation is exemplified and we see that FEM-C is relatively non-oscillatory by comparison. 

\begin{figure}[htbp]
\subfigure[Complete density profile]{
\label{subfig:sodshockCompareDiffBetaZoomOut}
\includegraphics[scale=0.27]{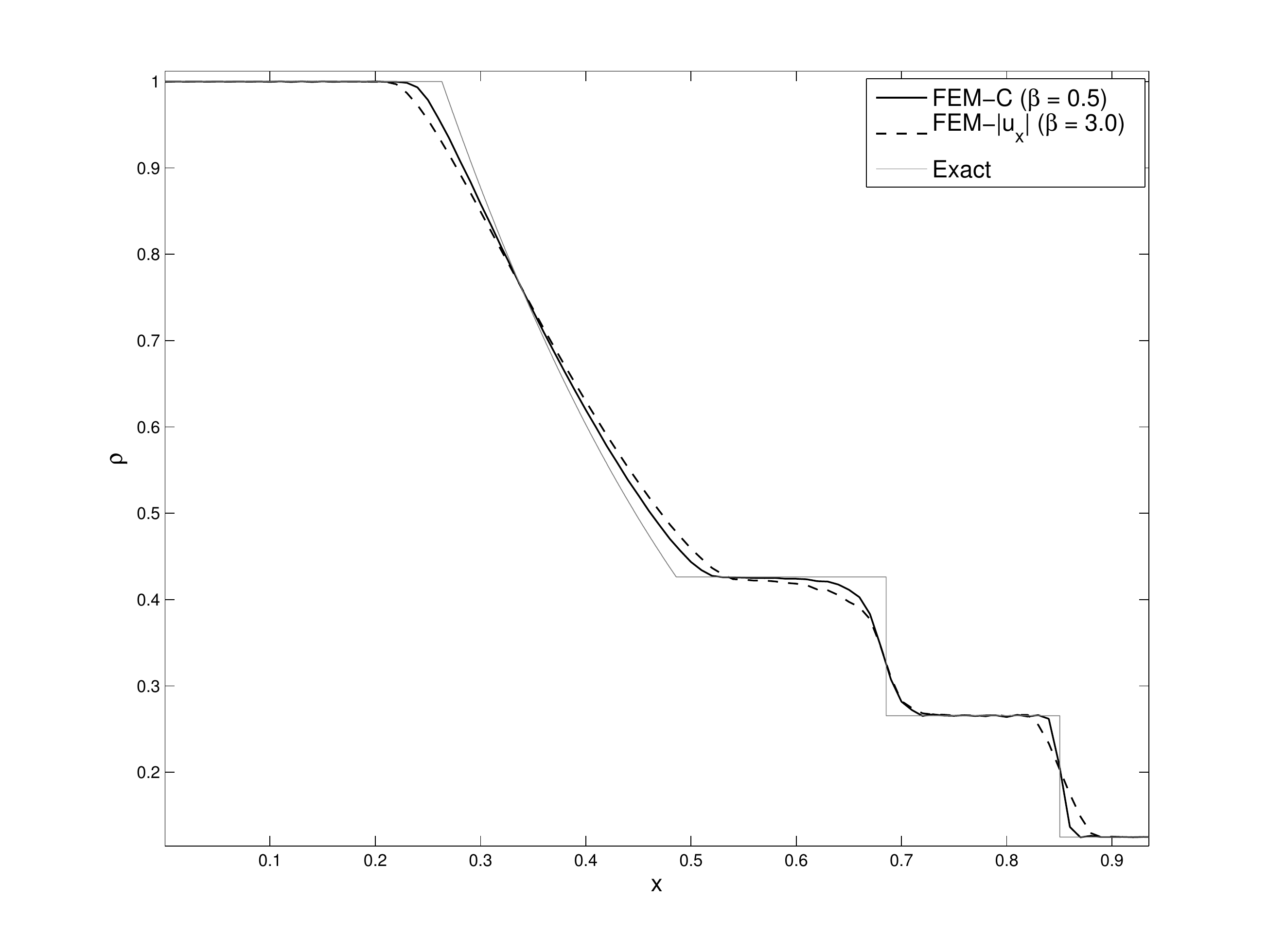}
}
\subfigure[Zoom-in at shock]{
\label{subfig:sodshockCompareDiffBetaZoomIn}
\includegraphics[scale=0.27]{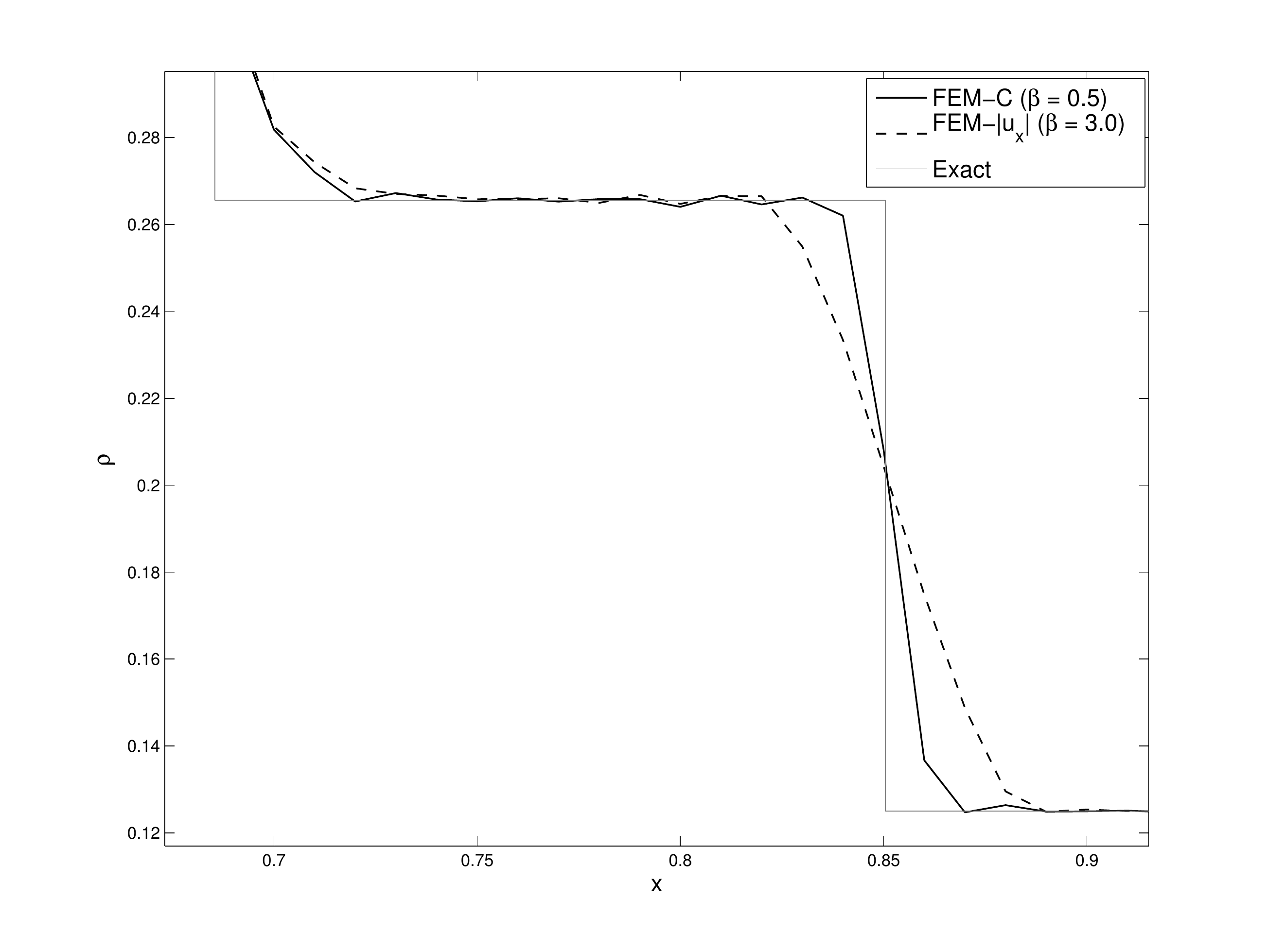}
}
\caption{Comparison of FEM-C and FEM-$|u_x|$, for the Sod shock-tube experiment with $N=100$, $T=0.2$. $\beta = 0.5$ for FEM-C and $\beta = 3.0$ for FEM-$|u_x|$.}
\label{fig:sodshockCompareDiffBeta}
\end{figure}

To limit these oscillations generated by FEM-$|u_x|$, we increase $\beta$ by a factor of $6$ and compare the resulting density in Figure \ref{fig:sodshockCompareDiffBeta}. In Figure \ref{subfig:sodshockCompareDiffBetaZoomIn} we can see a significant loss in accuracy when increasing to $\beta = 3$. Furthermore, in Figure \ref{subfig:sodshockCompareDiffBetaZoomOut}  we see FEM-$|u_x|$ requires  6 cells to capture the shock. 

\begin{figure}[htbp]
\subfigure[FEM-C vs. NT]{
\label{subfig:sodshockNT}
\includegraphics[scale=0.27]{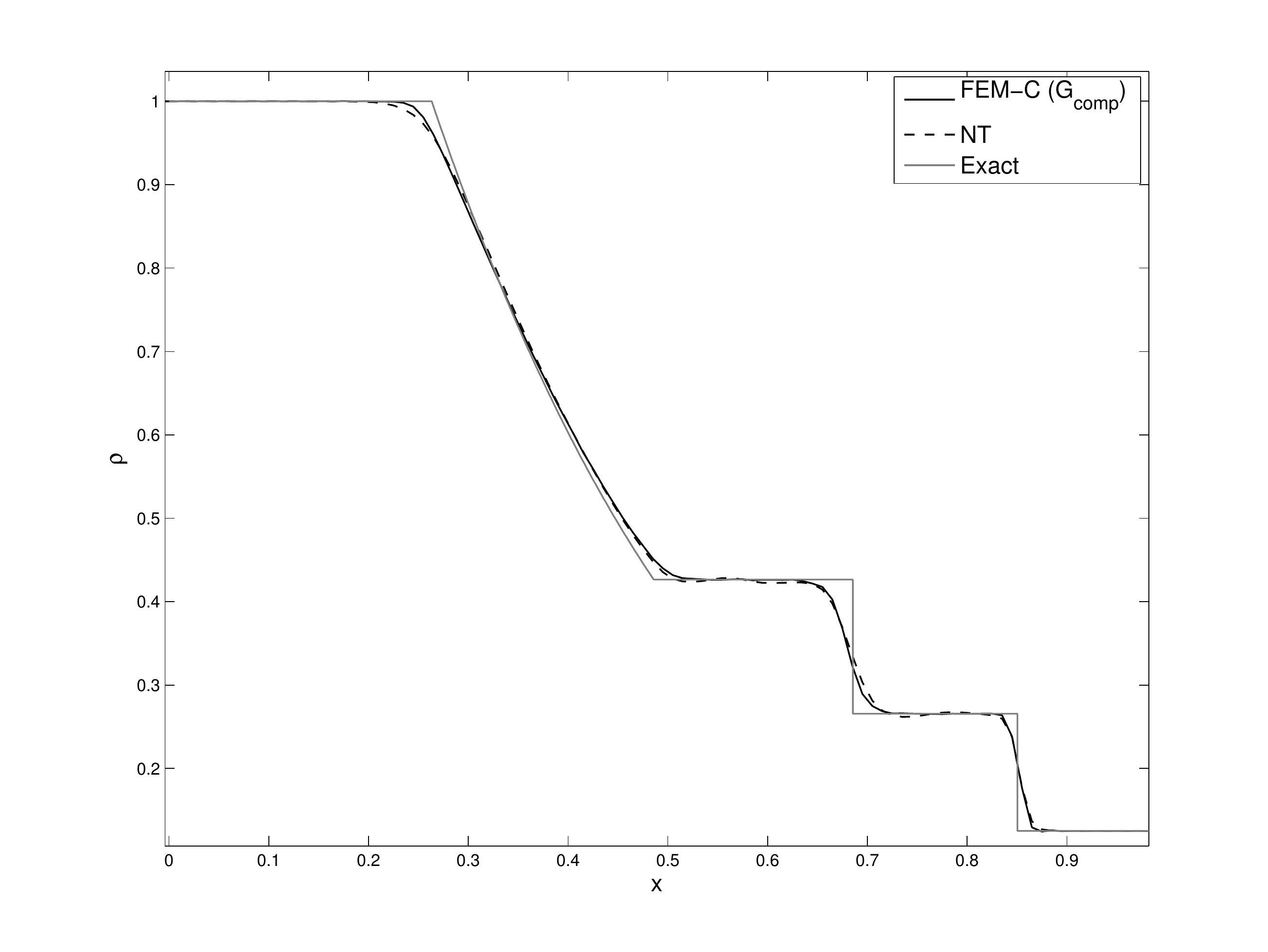}
}
\subfigure[FEM-C vs. WENO]{
\label{subfig:sodshockWENO}
\includegraphics[scale=0.27]{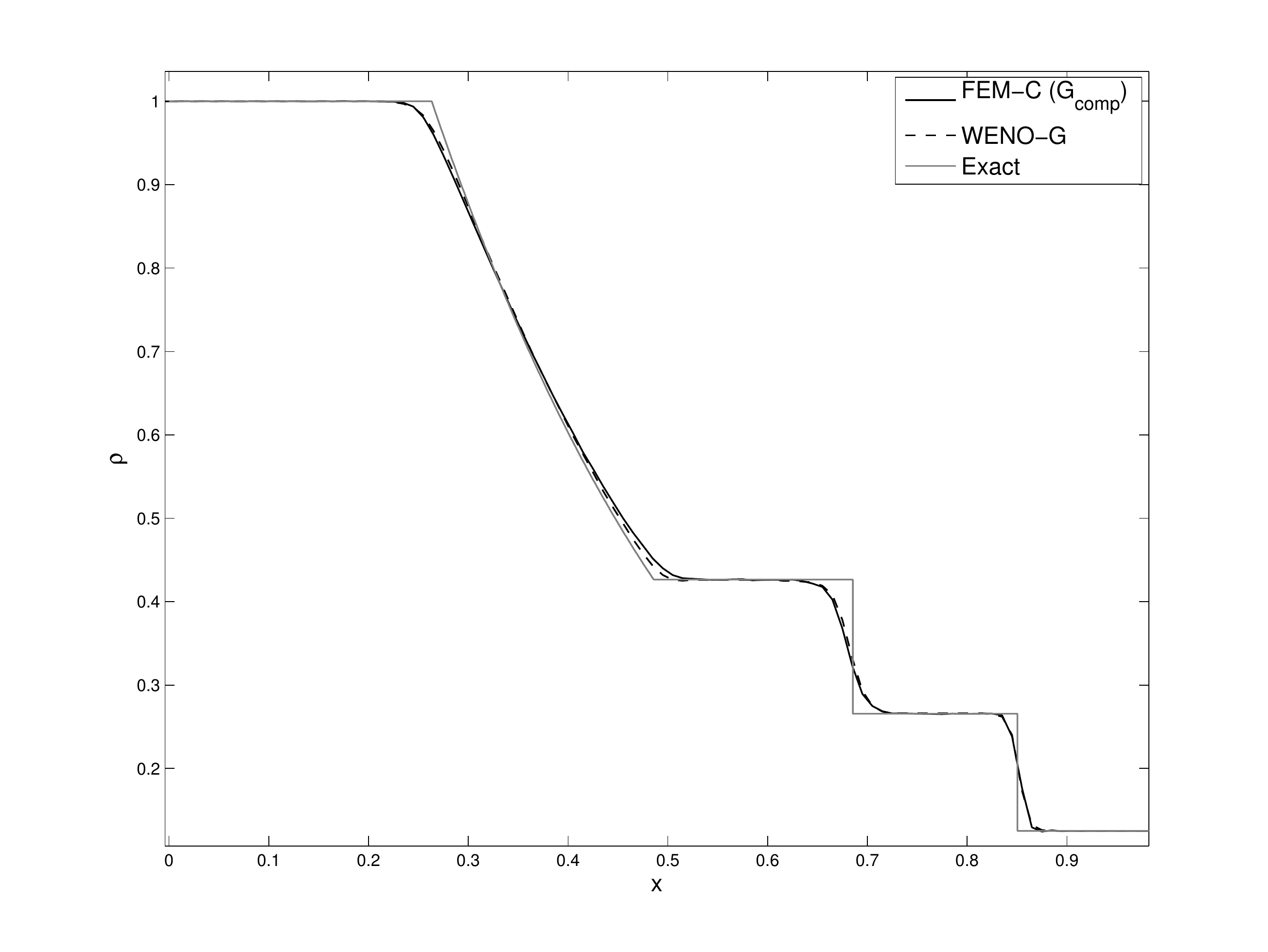}
}
\caption{Comparisons of FEM-C against NT and WENO schemes, for the Sod shock-tube experiment with $N=100$ and $T=0.2$.}
\label{fig:SSEFEMCompare}
\end{figure}

In Figure \ref{fig:SSEFEMCompare} we compare the results of the FEM-C scheme versus NT and WENO-G. Each simulation is performed with $N=100$ and for the FEM-C scheme we choose $\beta = 0.4$ and now use $G_{comp}$ (see Section \ref{subsec:modificationsForG}).

Each scheme produces similar resolution of the shock and contact discontinuity, capturing the shock in $3$ cells and the contact discontinuity in $6$ cells. The NT-scheme produces small, smooth, non-physical oscillations as the density transitions from the rarefaction to the lower states, and performs the worst at the rarefaction. Both FEM-C and WENO-G are essentially non-oscillatory and despite WENO-C performing slightly better at the rarefaction, the results are virtually indistinguishable at the shock and contact discontinuity. 

\section{Osher-Shu shock-tube problem}
\label{sec:osherShu}
For the problem of Osher-Shu, we consider the domain $\mathcal{I} = [-1,1]$ along with initial conditions
 \begin{equation}
 \label{eqn:osherShuIC}
 \left ( \begin{array}{c} \rho_0(x) \\ m_0(x) \\ E_0(x) \end{array} \right ) = \left ( \begin{array}{c} 3.857143 \\ 10.14185 \\ 39.1666 \end{array} \right )  {\bf 1}_{[-1,-0.8)} (x) +   \left ( \begin{array}{c} 1+0.2 \sin (5 \pi x) \\ 0 \\ 2.5 \end{array} \right ) {\bf 1}_{[-0.8,1]}(x),
 \end{equation}
  imposing natural boundary conditions at $x= -1$ and $x=1$

\vspace{10pt}

\begin{figure}[htbp]
\subfigure[FEM-C vs. NT]{
\label{subfig:osherShuNT}
\includegraphics[scale=0.27]{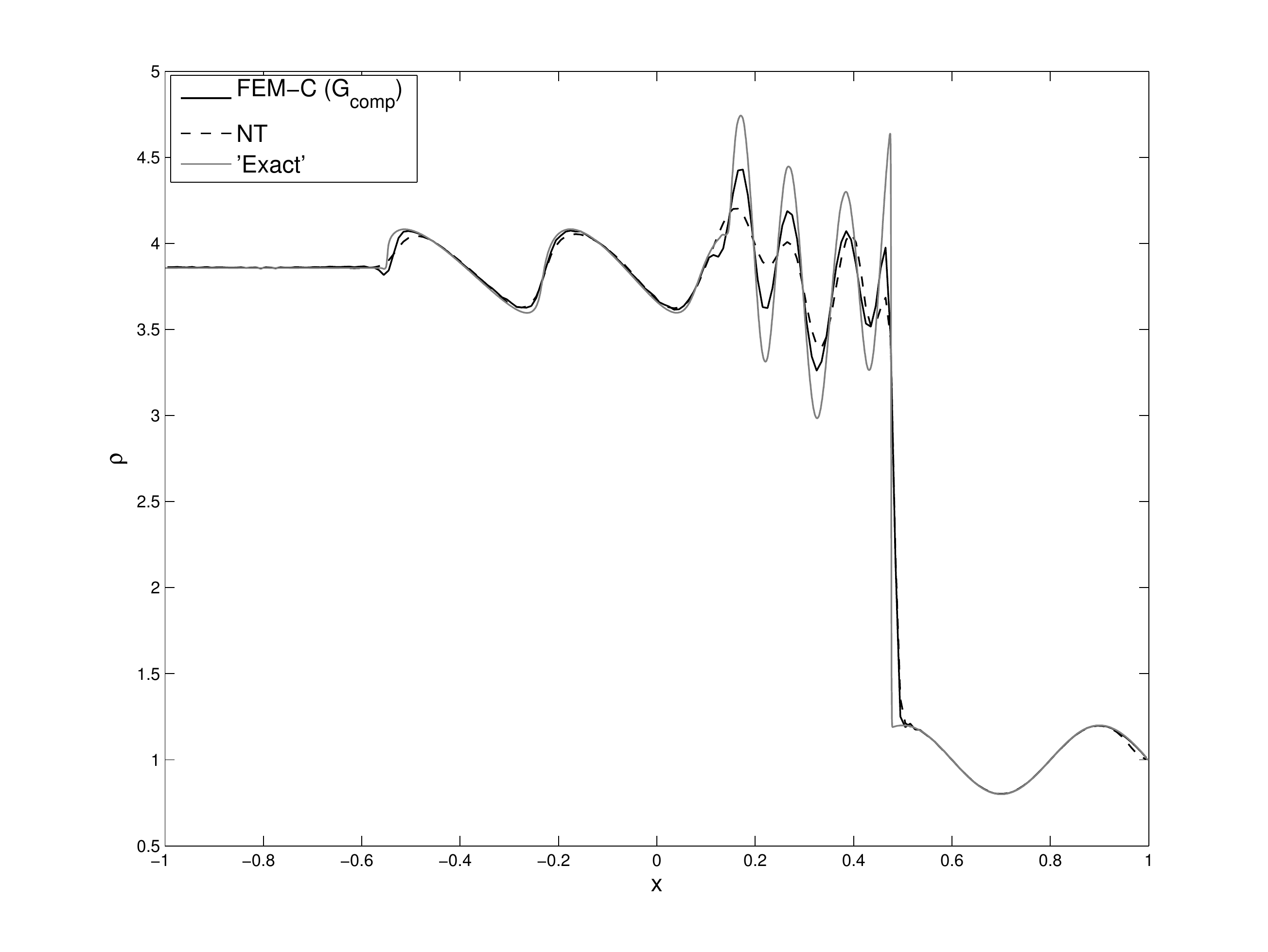}
}
\subfigure[FEM-C vs. WENO]{
\label{subfig:osherShuWENOGod}
\includegraphics[scale=0.27]{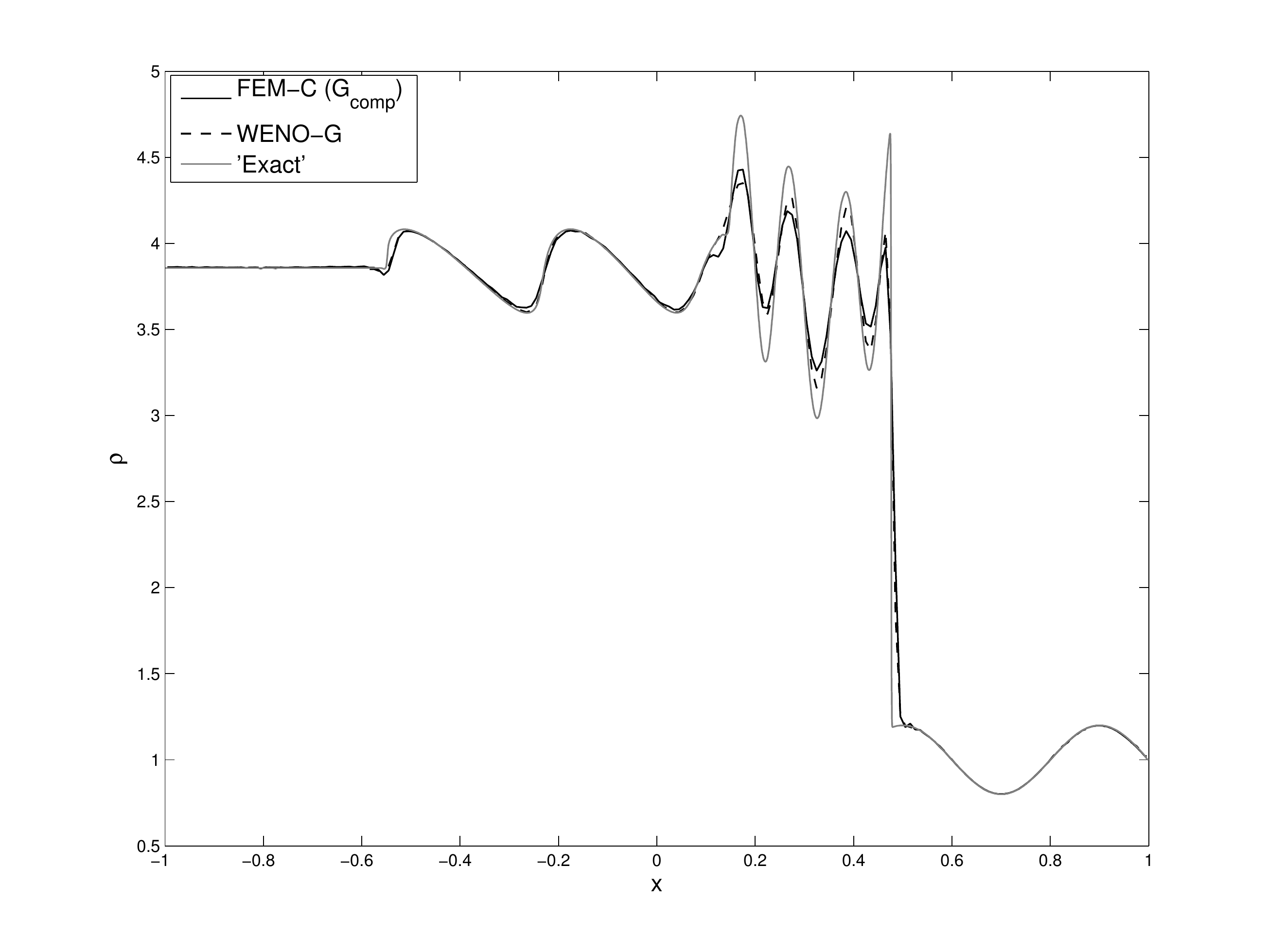}
}
\caption{Comparisons of FEM-C against NT and WENO schemes, for the  Osher-Shu shock-tube experiment with $N=200$ and $T=0.36$.}
\label{fig:OSEFEMCompare}
\end{figure}

This moderately difficult test problem, first considered in \cite{Shu198932}, proves to be more difficult for numerical schemes due to the evolution a shock-wave which interacts with an entropy-wave; care is required  to accurately capture the amplitudes of the post-shock entropy waves. Since the density is not monotone, standard flux limiters may unnecessarily apply too much dissipation at local-extrema, significantly reducing accuracy. An exact solution for this problem is not available and our `Exact' solution in our plots is generated using the DG-solver furnished in Hesthaven \& Warburton \cite{Hesthaven} with 3200 cells.

In Figure \ref{fig:OSEFEMCompare} we compare the results of FEM-C (we choose $\beta = 0.5$ and use $G_{comp}$), versus NT and WENO-G at $t=0.36$. In Figure \ref{subfig:osherShuNT} we see that NT diffuses the post-shock amplitudes and FEM-C provides far superior results. On the other hand, in Figure \ref{subfig:osherShuWENOGod} we see that all but one of the post-shock amplitudes are slightly better for the WENO-G scheme. This insufficiency of the FEM-C scheme is not completely surprising as the FEM-C is only formally second-order versus the fifth-order accuracy of the WENO-G scheme.

\begin{figure}[htbp]
\subfigure[WENO-C vs. WENO-G]{
\label{subfig:OSEWENOCWENOGodCompare}
\includegraphics[scale=0.27]{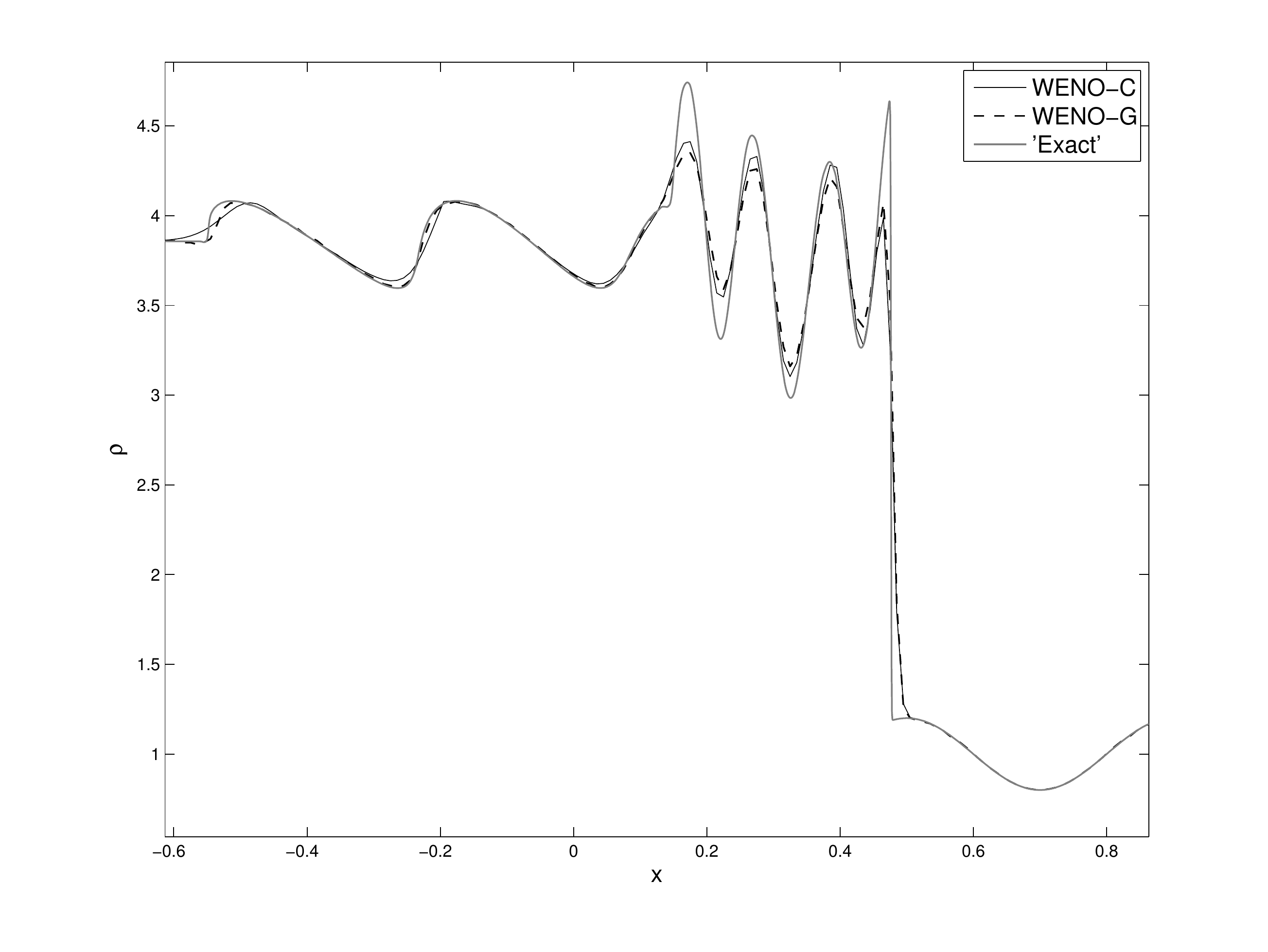}
}
\subfigure[WENO-C vs. WENO]{
\label{subfig:OSEWENOCWENOCompare}
\includegraphics[scale=0.27]{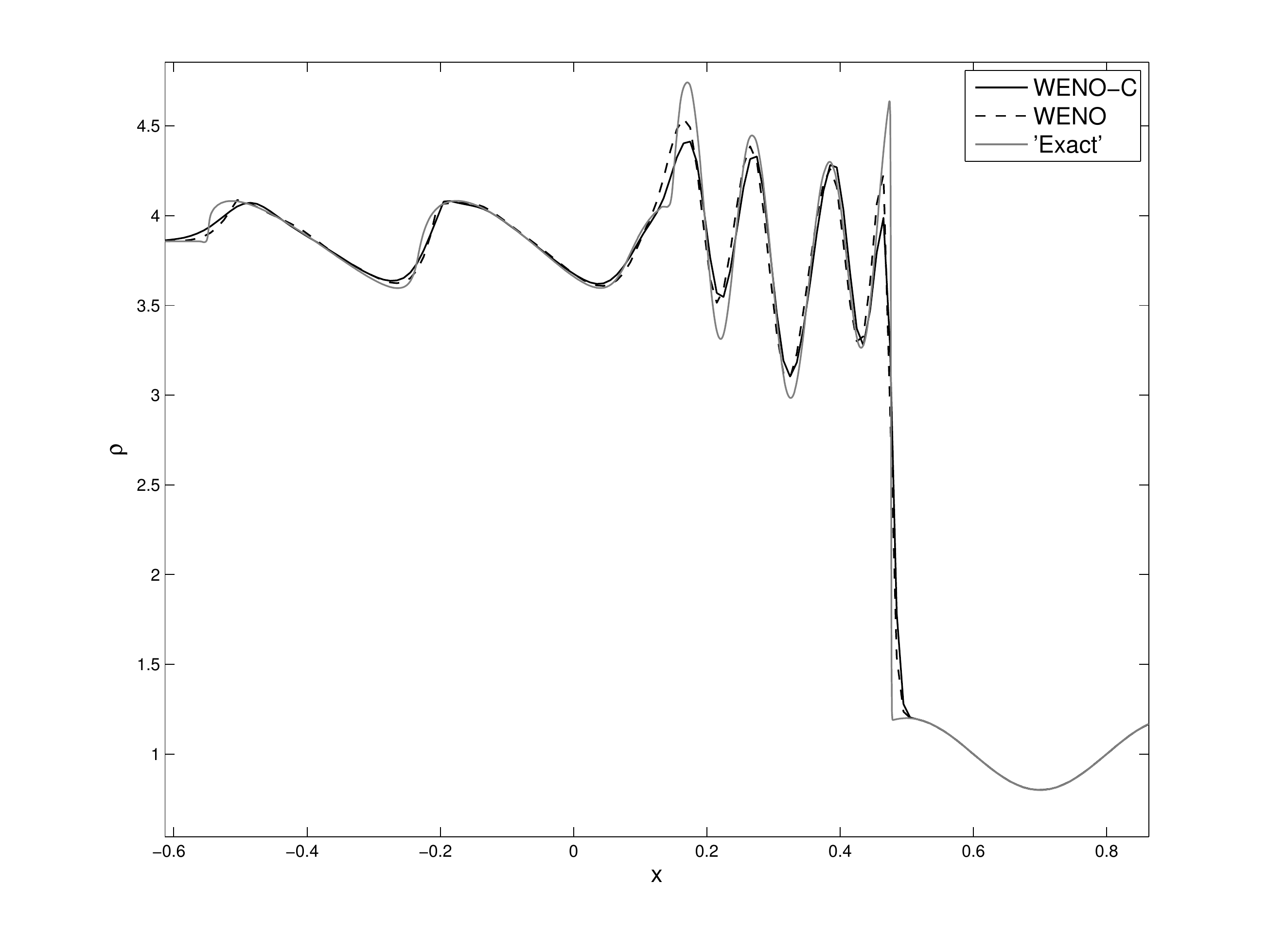}
}
\caption{Comparisons of WENO-C with WENO-G and our WENO scheme with artificial viscosity deactivated, for the Osher-Shu shock-tube experiment with $N=200$ and $T=0.36$.  }
\label{fig:OSEWENOCompare}
\end{figure}

Noting this insufficiency of the FEM-C scheme, we compare the WENO-G scheme with WENO-C in Figure \ref{subfig:OSEWENOCWENOGodCompare} and see the WENO-C scheme is more accurate in resolving the post-shock amplitudes. This comes at a price however, as we see WENO-G is more accurate in the N-wave region $[-0.6,0]$. 

Furthermore, it is interesting to note that in Figure \ref{subfig:OSEWENOCWENOCompare} where we choose $\beta = 0$ in our simplified WENO-scheme, we see that the C-equation is not necessary for Osher-Shu. As we see in Section \ref{sec:woodwardColella} this ceases to be the case as the collision of strong shock waves require stabilization. 

\section{Woodward-Colella Blast Wave}
\label{sec:woodwardColella}

The colliding blast wave problem of Woodward-Collella is posed on the domain $\mathcal{I} = [0,1]$ with initial conditions
 \begin{gather*}
\rho_0(x) = 1, \\ m_0(x) = 0, \\ 
E_0(x)= 250 \cdot {\bf 1}_{[0.9,1]} + 0.25 \cdot  {\bf 1}_{[0.1,0.9)} + 2500 \cdot {\bf 1}_{[0,0.1)},
\end{gather*}
and reflective boundary conditions at $x=0$ and $x=1$. This challenging blast wave problem, considered in
\cite{Colella1984115} tests the ability of a numerical scheme to handle collisions between strong shock waves. Any viable scheme generally requires stabilization at these collisions. For the results of a wide range of schemes applied to this problem,  see \cite{Colella1984174}.  An exact solution for this problem is not available and the `Exact' solution in our plots is generated with a 400-cell PPM solver.

\begin{figure}[h!tbp]
\subfigure[FEM-C vs. NT]{
\includegraphics[scale=0.27]{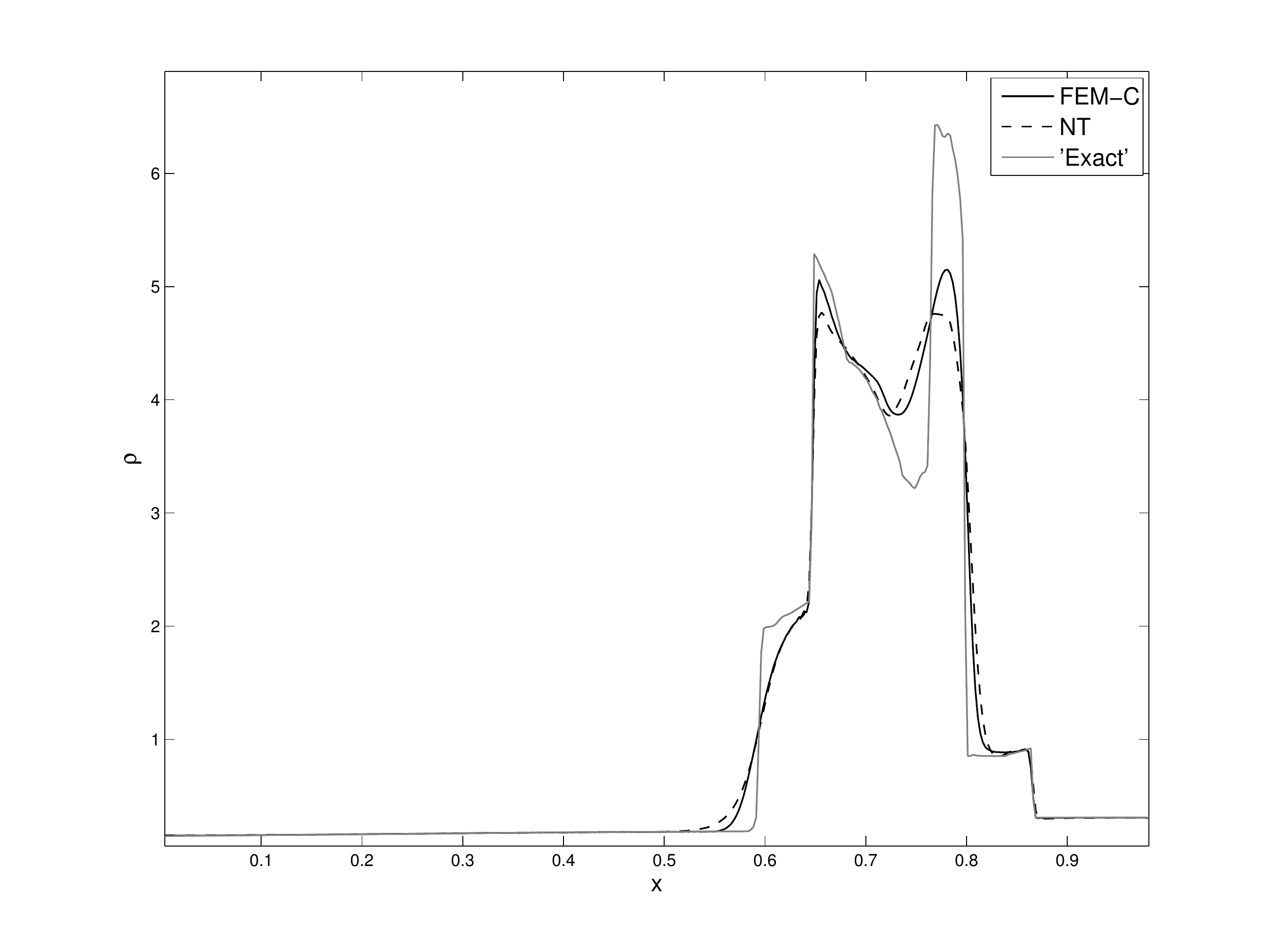}
}
\subfigure[FEM-C vs. WENO]{
\includegraphics[scale=0.27]{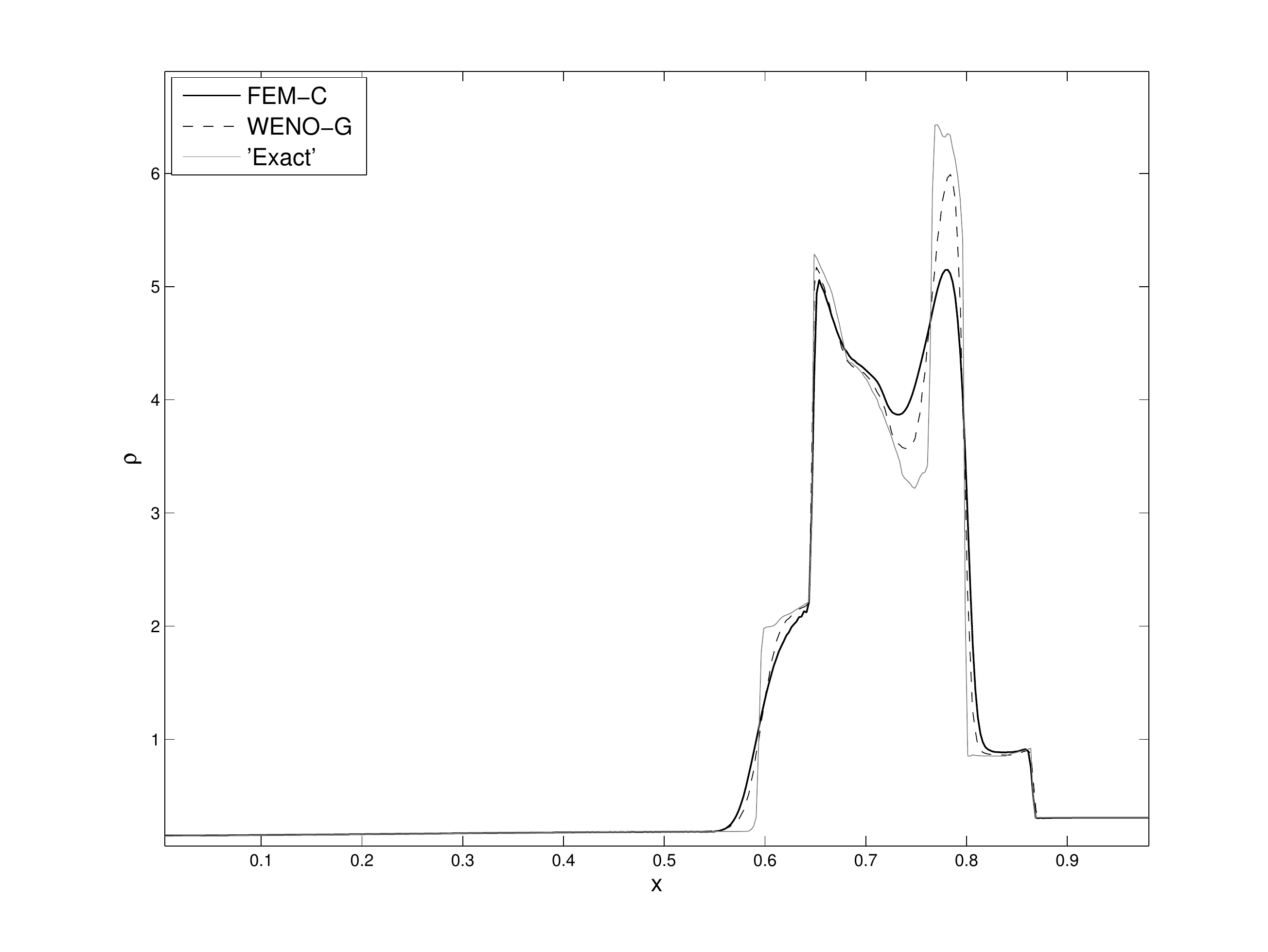}
}
\caption{Comparisons of FEM-C against NT and WENO schemes, for the Woodward-Colella blast-tube experiment with $N=400$ and $T = 0.038$.}
\label{fig:WCEFEMCompare}
\end{figure}

As is standard in our sequence of experiments,  we provide a comparison of FEM-C ($\beta = 0.5$) with NT and WENO-G in Figure \ref{fig:WCEFEMCompare} at $t=0.038$. It is interesting to note, the use of $G_{comp}$ is far too oscillatory in this difficult test problem; we revert to the standard choice of $G$. We again see that while FEM-C is superior to NT in capturing the amplitude of the two peaks in the density, FEM-C is far too diffusive in comparison to WENO-G.

Despite the relative inefficiency of FEM-C compared to WENO-G, it is interesting to note that our FEM-C results (with $N=1200$) are better than the artificial viscosity schemes use in
Colella \& Woodward
\cite{Colella1984174}. Our scheme is slightly sharper at the shocks and contact discontinuities and is just as accurate in the height of the two peaks.

\begin{figure}[h!tbp]
\subfigure[WENO (without stabilization)]{
\includegraphics[scale=0.27]{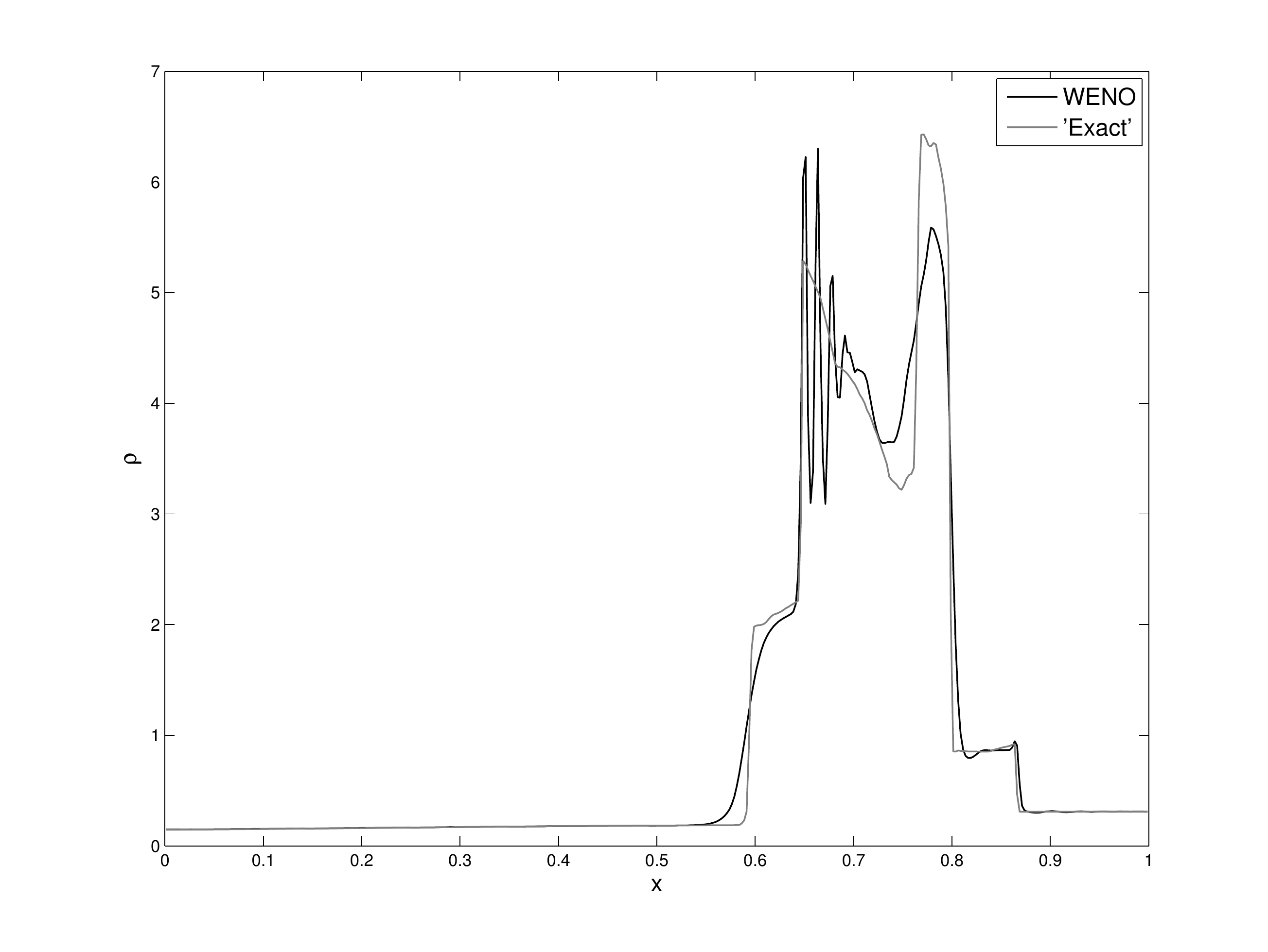}
\label{subfig:WENONoStab}
}
\subfigure[WENO (with stabilization)]{
\includegraphics[scale=0.27]{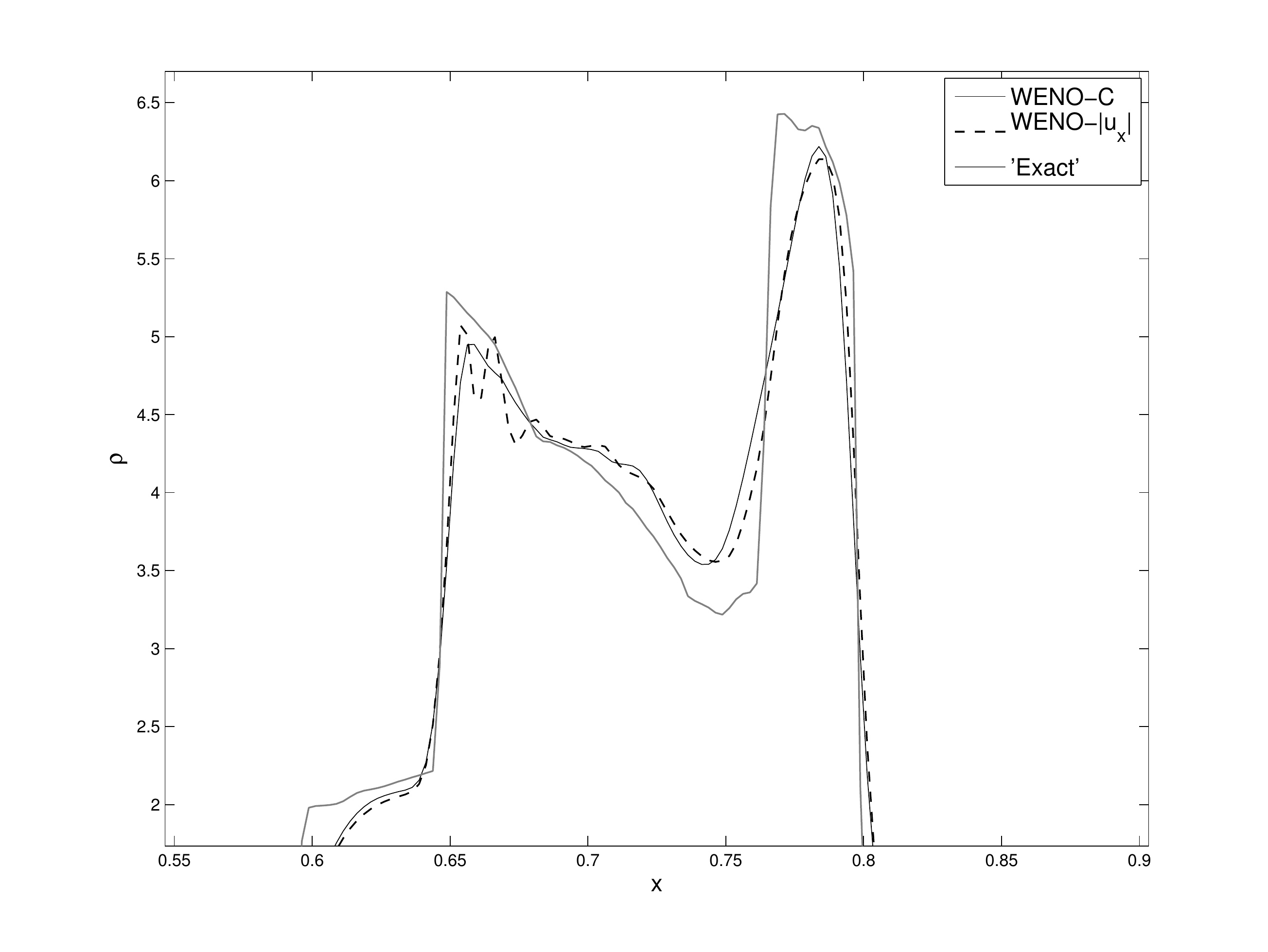}
\label{subfig:WENOStab}
}
\caption{WENO with and without stabilization applied to the Woodward-Colella blast-tube experiment with $N=400$ and $T=0.038$.}
\label{fig:WCEWENOExactCompare}
\end{figure}

Before moving to a comparison of WENO-G and WENO-C, in Figure \ref{subfig:WENONoStab} we see that our simplified WENO scheme is highly oscillatory due to the strong shock collision, necessitating the use of stabilization. This requirement contrasts to the observations made in Section \ref{sec:osherShu}. However,  in Figure \ref{subfig:WENOStab}, we see that the use of a classical artificial viscosity significantly dampens the instability but moderate oscillations occur and the $C$-method provides similar dampening in a smooth way.

\begin{figure}[htbp]
\begin{center}
\includegraphics[scale=0.3]{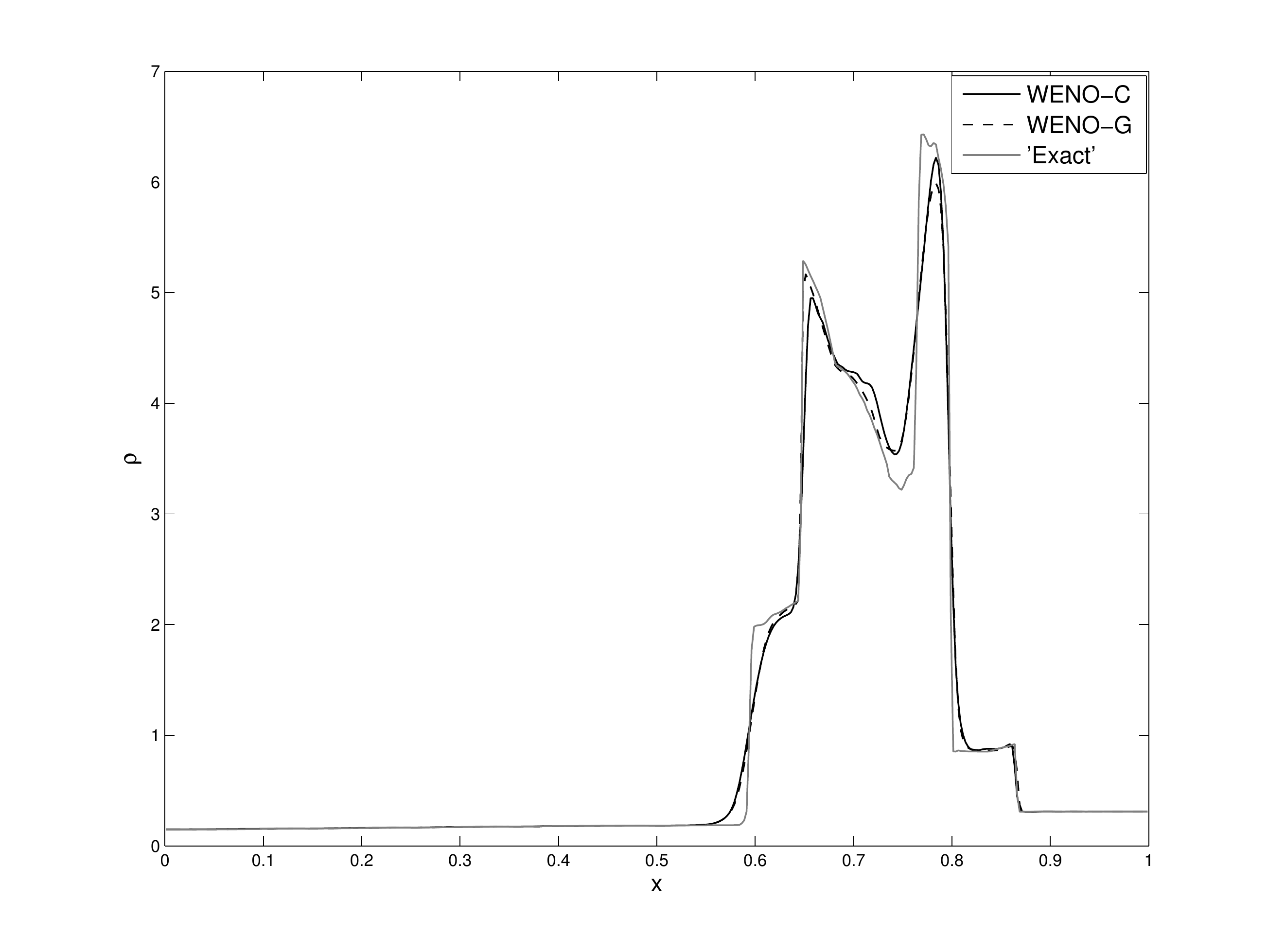}
\caption{Comparison of WENO-C against WENO-G,  for the Woodward-Colella blast-tube experiment with $N=400$ and $T=0.038$.}
\label{fig:WCEWENOCWENOGodCompare}
\end{center}
\end{figure}

Finally, in Figure \ref{fig:WCEWENOCWENOGodCompare} we demonstrate the relative success of WENO-C versus WENO-G. At the left peak, WENO-G is more accurate, but at the right peak the reverse situation occurs. Each scheme provides very good results, and it is clear that WENO-C is a simple alternative to WENO-G which produces similar results for complicated shock interaction. 

\section{Leblanc shock-tube problem}
\label{sec:leblanc}
We conclude our experiments with the Leblanc shock-tube, posed on the domain $\mathcal{I} = [0,9]$, with initial conditions

\begin{equation}
\label{eqn:leblancIC}
\left ( \begin{array}{c} \rho_0(x) \\ m_0(x) \\ E_0(x) \end{array} \right ) = 
\left ( \begin{array}{c} 1 \\ 0 \\ 10^{-1} \end{array} \right ) {\bf 1}_{[0,3)}(x) + 
\left ( \begin{array}{c} 10^{-3} \\ 0 \\ 10^{-9} \end{array} \right ) {\bf 1}_{ [3,9]} (x)\,,
\end{equation}
 with natural boundary conditions at $x=0$ and $x=9$, and with the adiabatic constant $\gamma = \frac{5}{3}$. 

Because the initial energy $E_0$ jumps ten orders of magnitude, a very strong shock wave is produced, making the Leblanc problem  an extraordinarily difficult numerical experiment.
First , numerical methods tend to over-estimate the correct  shock speed whenever the shock wave in the pressure field is not sharply resolved.   
Second, numerical  approximations tend to produce  large overshoots in the internal energy
\[
e = \frac{p}{(\gamma - 1)\rho}
\]
 at the contact discontinuity.   We refer the reader to  Liu, Cheng, \& Shu  \cite{Liu20098872} and
 Loub\'{e}re \& Shashkov  \cite{Loubere2005105} for a discussion of the difficulties in the numerical simulation of the Leblanc problem for a variety of
 numerical schemes.   The second-order finite-element basis that we use for our FEM-C algorithm is not sufficiently high-order to accurately capture
 wave speeds in Leblanc, but our  fifth-order WENO-C scheme  is ideally suited for this difficult test case.   We shall present two differing strategies for WENO-C,
 which both capture the correct shock speed and remove overshoots of the internal energy.

\subsection{Strategy One: A $C$ equation for the energy density}   As we introduced the $C$-method in equation  (\ref{subeq:cmethodEuler}), artificial viscosity is present on
the right-hand side of all three conservation laws for momentum, mass, and energy.   For the WENO-C algorithm, only viscosity in the momentum equation has been used for
the Sod, Osher-Shu, and Woodward-Colella test cases.  Due to the strength of the shock in Leblanc, we now return to using artificial viscosity for the energy equation.
In our first strategy for this problem, we solve for one additional linear reaction-difffusion equation for a new $C$-coefficient to use on the right-hand side of the energy conservation law.

Specifically, to combat the large overshoot in the internal energy $e$, we solve a second $C$-equation for the coefficient which we label $C_E$; the forcing term for the
$C_E$ equation uses $| \partial_x (E/\rho)|/ \max |\partial_x(E/\rho)|$, replacing $|\partial_x u|/\max |\partial_x u|$ which forces the $C$-equation for the coefficient
$C_u$, used for the right-hand side of the momentum equation.   In particular, since $C_u$ is found using  the $G_{comp}$ forcing, activated only in compressive regions when $u_x <0$,
for the $C_E$ equation, we activate the right-hand side only in expansive regions when $u_x \ge 0$.
To be precise, this modified WENO-C scheme replaces the semi-discrete form \eqref{eqn:wenoCSemiDiscrete} with 
\begin{equation}
\label{eqn:wenoCModSemiDiscrete}
\partial_t \left [ \begin{array}{c} {\bf u}_i \\ {\bf C}_i \end{array} \right ] + \frac{1}{\Delta x} \left [\begin{array}{c} \mathcal{\tilde A}_{\text{WENO}} ({\bf u}_i, {\bf C}_i) \\ \mathcal{\tilde B}_{\text{WENO}} ({\bf u}_i, {\bf C}_i) \end{array} \right ] = 0.
\end{equation}
The resulting fully-discrete scheme solves for ${\bf u}^n_i$ and
\[
{\bf C}^n_i = \left ( \begin{array}{c} C^n_{u_i} \\ C^n_{E_i} \end{array} \right )
\]
where the modified fluxes $\mathcal{\tilde A}_{\text{WENO}}$ and $\mathcal{\tilde B}_{\text{WENO}}$ are given by:
\begin{subequations}
\begin{multline}
\left [ \mathcal{\tilde A}_{\text{WENO}}  \left ( \left [ \begin{array}{c} \rho_i \\ m_i \\ E_i  \end{array} \right ], \left [ \begin{array}{c} C_{u_i}  \\ C_{E_i}  \end{array} \right ] \right ) \right ] = \\ \left  [\begin{array}{c} \text{WENO}(\rho_i, u_{i \pm 1/2 } ) \\ 
\text{WENO}(m_i , u_{i \pm 1/2} ) + \tilde \partial p_i - { \frac{ \tilde \partial_{C_u} u_{i+1/2} - \tilde \partial_{C_u} u_{i-1/2}}{\Delta x} } \\
\text{WENO}_E(E_i, u_{i \pm 1/2} )  - { \frac{ \tilde \partial_{C_E} E_{i+1/2} - \tilde \partial_{C_E}  E_{i-1/2}}{\Delta x} } \end{array}
\right ]
\end{multline}
\begin{multline}
\left [ \mathcal{\tilde B}_{\text{WENO}} \left ( \left [ \begin{array}{c} \rho_i \\ m_i \\ E_i  \end{array} \right ], \left [ \begin{array}{c} C_{u_i} \\ C_{E_i} \end{array} \right ] \right ) \right ] = \\  \frac{1}{\Delta x} \left [ \begin{array}{c} - {S({\bf u}_i)} \left [C_{u_i} - G_{comp}(\tilde \partial u_i) \right ] + {\tilde \partial_S C_{u_{i+1/2}} + \tilde \partial_S C_{u_{i-1/2}} } \\
- {S({\bf u}_i)}\left [C_{E_i} - G_{expand}(\tilde \partial (E / \rho)_i, \tilde \partial u_i) \right ] +  {\tilde \partial_S C_{E_{i+1/2}} - \tilde \partial_S C_{E_{i-1/2}} }.
\end{array} \right ]
\end{multline}
\end{subequations}
The expansive-region forcing for $C_E$ is given by
\begin{equation}
\label{eqn:Gexpand}
G_{expand}\left ( \tilde \partial E_i, \tilde \partial u_i \right ) = \frac{|\tilde \partial (E / \rho)_i |}{\underset{i}{\max}  |  \tilde \partial (E / \rho)_i |} {\bf 1}_{[0,\infty)}(\tilde \partial u_i)
\end{equation}
and we use the shorthand
\[
\tilde \partial_{C_u} u_{i+1/2} = \beta_u \ \Delta x^2 \ \max_{i} \left | \tilde \partial u_{i+1/2} \right | \  \frac{C_{u_{i+1/2}}}{\underset{i}{\max}  \ C_{u_i}} \ \rho_{i+1/2} \  \tilde \partial u_{i+1/2},
\]
and
\[
\tilde \partial_{C_E} E_{i+1/2} = \beta_E \ \Delta x^2 \ \max_{i} \left | \tilde \partial u_{i+1/2} \right | \  \frac{C_{E_{i+1/2}}}{\underset{i}{\max}  \ C_{E_i}} \rho_{i+1/2} \ \tilde \partial (E / \rho)_{i+1/2}.
\]

\begin{figure}[htbp]
\subfigure[With and without $C_{E}$, $N=1440$]{
\includegraphics[scale=0.27]{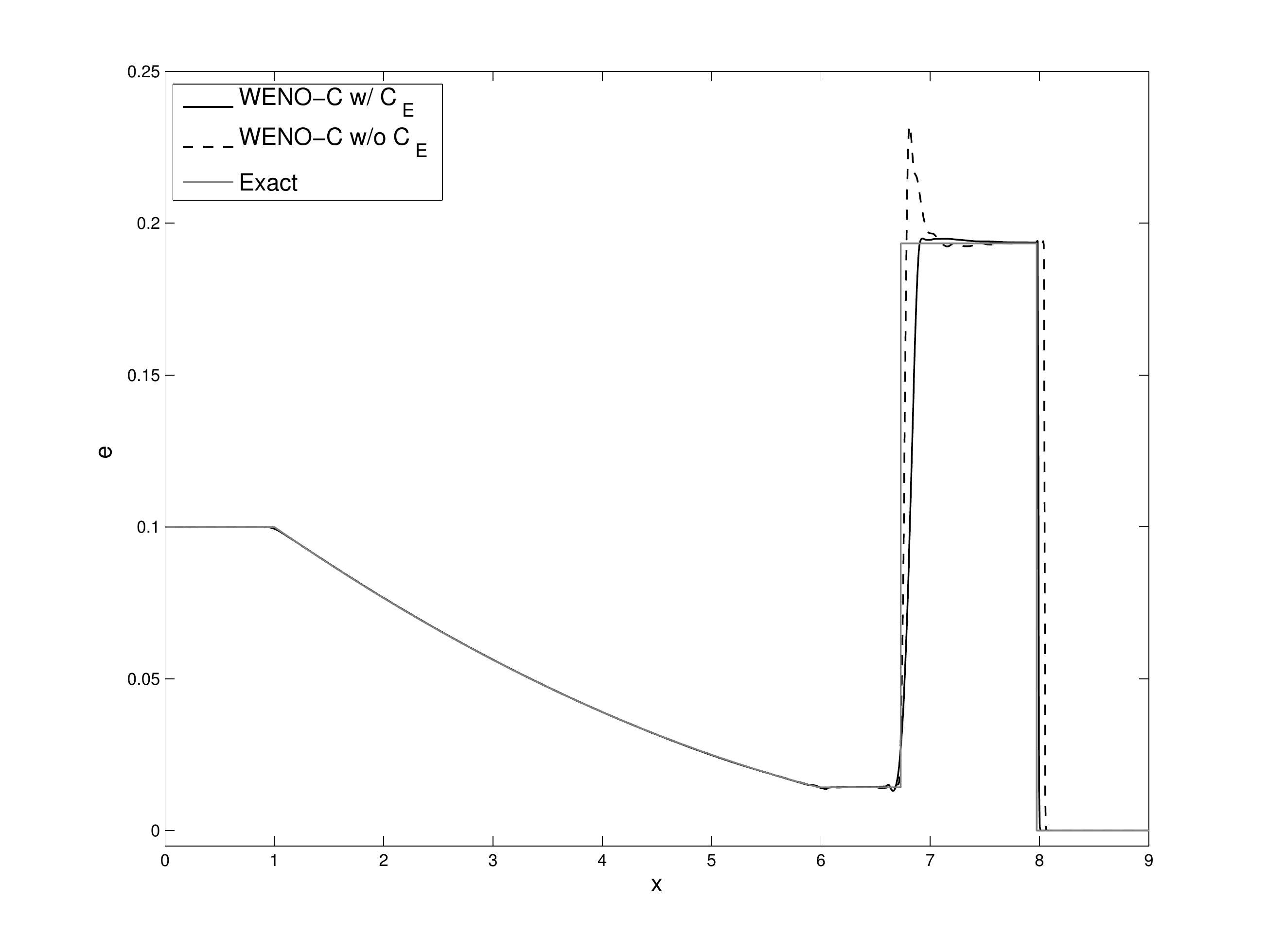}
\label{subfig:LECEComparison}
}
\subfigure[Successive refinements, $N=360,720,1440$]{
\includegraphics[scale=0.27]{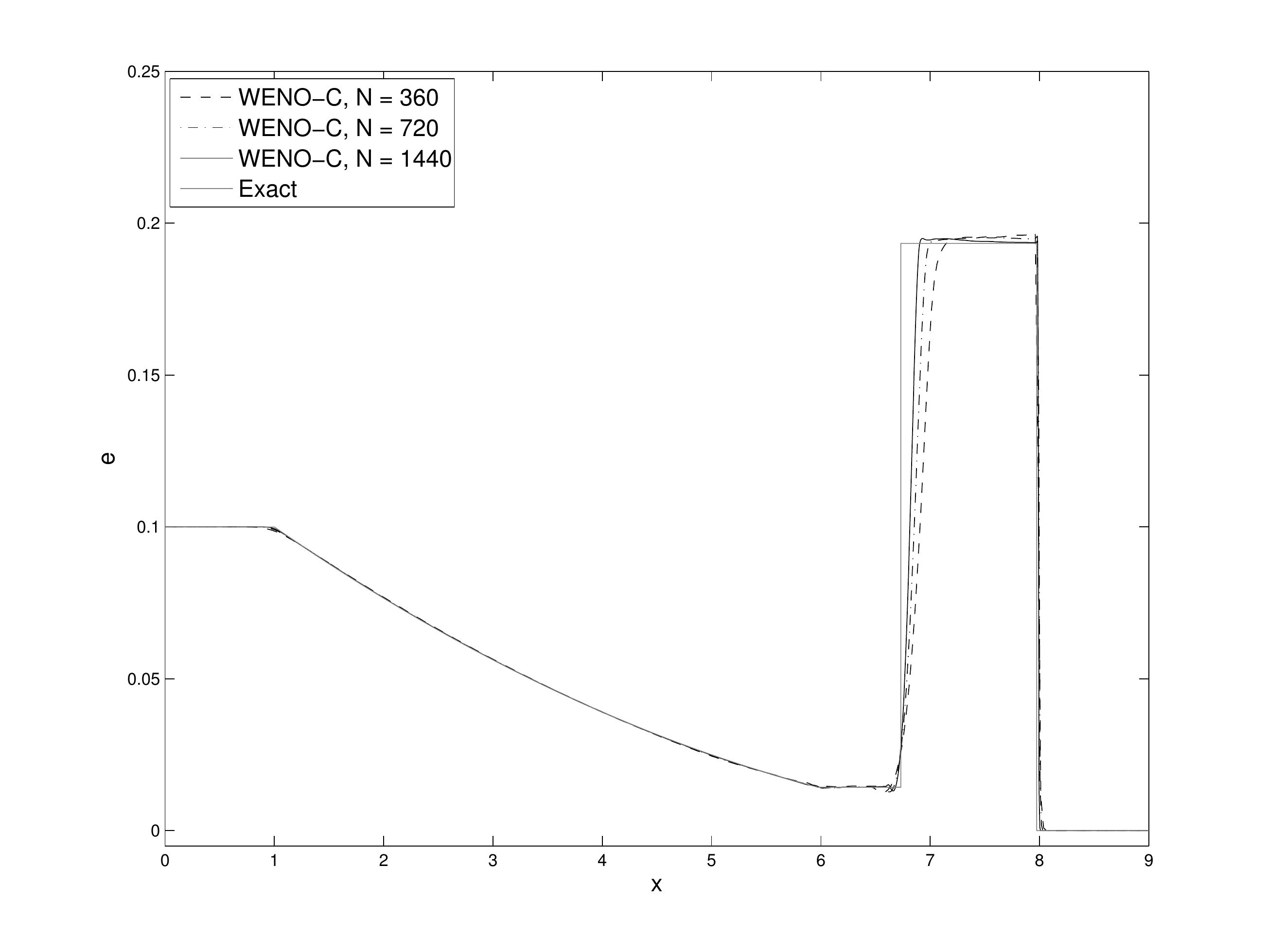}
\label{subfig:LEWENOCConvergence}
}
\caption{Internal energy plots for WENO-C for the Leblanc shock-tube experiment at $T=6$.}
\label{fig:LEWENO}
\end{figure}

In Figure \ref{subfig:LECEComparison} we plot the difference between WENO-C with and without the use of this new equation for $C_E$. For WENO-C with $C_E$ activated, we choose $\beta_u = 1.0$ and $\beta_E = 0.15$; with the $C_E$-equation deactivated,  we use $\beta_u = 1.0$ and $\beta_E = 0$.   Observe that activating the  $C_E$-equation  removes the large overshoot at the contact discontinuity. Furthermore, examining the location of the shock, we see that the use of the $C_E$-equation produces  more accurate approximations of  the shock speed. 

In Figure \ref{subfig:LEWENOCConvergence} we show the results of WENO-C at $N=360,720,1440$. In this plot, we see very little overshoot at each level of refinement and this small overshoot does not grow with refinement. 
 
\subsection{Strategy Two: a new type of viscosity for the energy density}
Our second strategy for the Leblanc problem may be viewed as being motivated by the  energy dissipation rate of real fluids, and adheres to our framework of only solving one $C$-equation, forced by the normalized modulus of
the gradient of velocity.    The idea is easy to explain, and we begin by writing the equations for momentum and mass (we drop the superscript $ \epsilon $):
\begin{subequations}
\label{leblancC}
\begin{align} 
 (\rho u)_t +  ( \rho u^2 + p)_x & = \epsilon^2 \tilde \beta  ( C \rho  u_x)_x \,, \\
\rho_t + (\rho u)_x &=0 \,\\
p&= (\gamma - 1)\rho \, e \,, \\
C_t - S(u) C_{xx} + \frac{S(u)}{ \epsilon } C  & = S(u) G(u_x) \,.
\end{align}
\end{subequations} 
By multiplying the momentum equation by the velocity $u$, integrating over the spatial domain, and using the conservation of mass equation, we find the
basic energy law:
\begin{equation}\label{energylaw2}
  \frac{d}{dt} \left[ \int  {\frac{1}{2}} \rho  u^2\, dx + {\frac{1}{\gamma -1}} \int p \, dx\right]  = -  \epsilon ^2\tilde\beta \int  C \rho  \,  u _x^2 \, dx \,.
\end{equation} 
Note, that when $ \epsilon =0$, the variable $E$ is exactly the energy density; that is, when $  \epsilon =0$, $E= {\frac{1}{2}} \rho  u^2 + {\frac{p}{\gamma -1}}$.   Thus,
we formulate a right-hand side term for the energy equation to ensure the $E$ continues to represent the energy density for $ \epsilon >0$.   To do,  we choose a right-hand side
which will provide the same energy law as (\ref{energylaw2}).  We introduce the following equation:
\begin{equation}\label{energy2}
E_t + (uE +up)_x = - \epsilon^2 \tilde \beta   C \rho  \,  u _x^2 \,.
\end{equation} 
The fundamental theorem of calculus shows that integration of (\ref{energy2}) provides the same basic energy law as (\ref{energylaw2}).   Hence, our second strategy employs the equation
(\ref{leblancC}) together with (\ref{energy2}).  The interesting feature of the new right-hand side of the energy equation is its nonlinear structure, quadratic in velocity gradients.  This energy
loss compensates for entropy production, and can become  anti-diffusive near contact discontinuities.     As such, we shall discretize this set of equations using the very stable Lax-Friedrichs
flux.      We remark that  the term $\epsilon^2 \tilde \beta   C \rho  \,  u _x^2$ is analogous to the
viscous dissipation term of the Navier-Stokes-Fourier system and can be found as a truncation error in \cite{GeMaDa1966}.

As we noted above,
to the best of our knowledge, the most commonly used numerical schemes applied to Leblanc tend to exhibit a significant overshoot in the internal energy $e$ at the contact discontinuity. 
Furthermore, on coarse meshes ($<2000$ cells), the speed of the shock tends to be inaccurate. Indeed, this is the case for arguably the most widely used WENO implementation,  designated 
WENO-LF-5-RK-4 by Jiang \& Shu \cite{Jiang1996202}. This scheme, which we call WENO-LF, uses a Lax-Friedrichs flux-splitting with a $5$th-order WENO reconstruction in space and $4$th-order 
Runge-Kutta in time.

\begin{figure}[htbp]
\subfigure[WENO-LF]{
\includegraphics[scale=0.27]{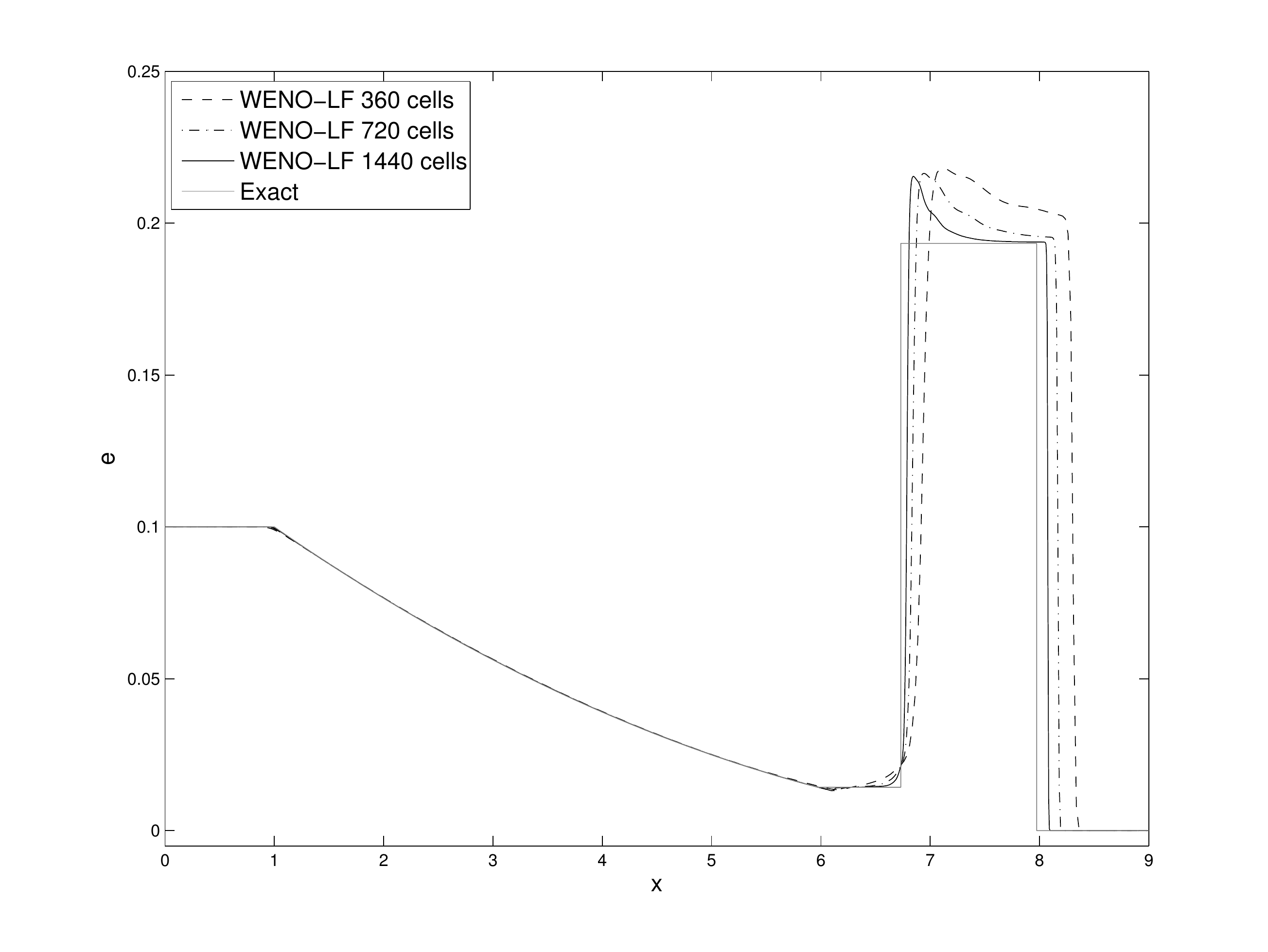}
\label{subfig:LBLFConverg}
}
\subfigure[WENO-LF-C]{
\includegraphics[scale=0.27]{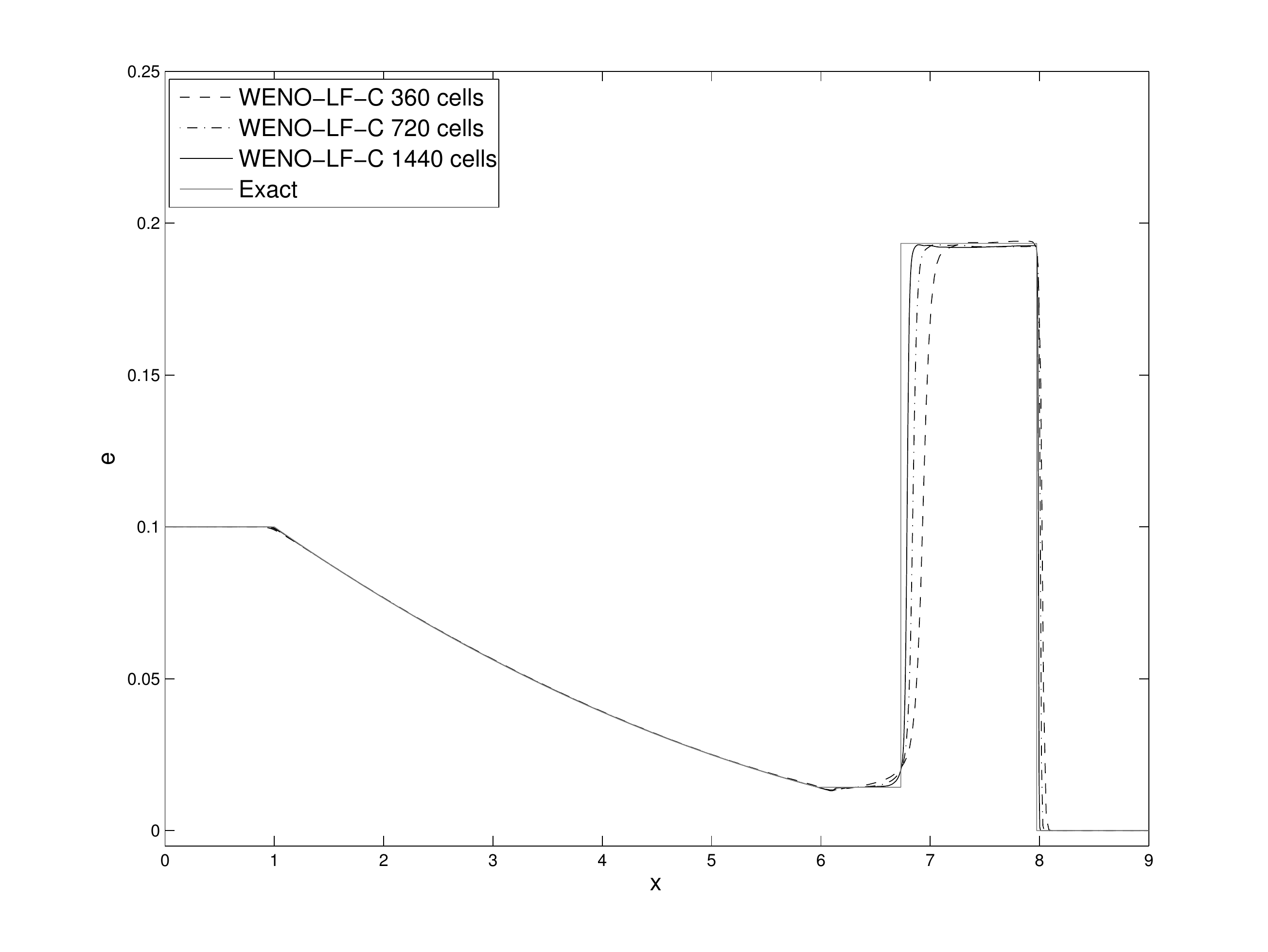}
\label{subfig:LBLFbeta5Converg}
}
\caption{Internal energy plots for the Leblanc shock-tube experiment  at $T=6$ using WENO-LF with and without the $C$-equation.}
\label{fig:LBLFWENO}
\end{figure}

If we examine the contact discontinuity at $x \approx 6.8$ in Figure \ref{subfig:LBLFConverg}, at resolutions $N=360,720,1440$ we see that WENO-LF exhibits relative overshoots of  $12.8 \%$, $11.8\%$ and $11.4\%$ respectively. This slow decay of the overshoot suggests that WENO-LF suffers from the Gibbs-phenomenon, despite it's attempt to quell oscillatory behavior. Examining the shock at $x\approx 8$ we see that the computed shock speeds are  inaccurate. 

To address the loss of accuracy exhibited by WENO-LF, we propose the use of the $C$-equation along with a nonlinear viscosity on the energy equation. Since WENO-LF has an intrinsic artificial viscosity (by virtue of the Lax-Friedrichs splitting) on the right-hand side of the momentum equation, we find that we do not need to explicitly use our artificial viscosity for the momentum (even though this
mathematically motivated our nonlinear viscosity for the energy equation).   As such, we 
 require a single $C$-equation which is forced by $G_{comp}(u_x)$.

Keeping consistent with the semi-discrete formulation, we write the WENO-LF-C scheme 
\begin{equation}
\label{eqn:wenoLFCSemiDiscrete}
\partial_t \left [ \begin{array}{c} {\bf u}_i \\ C_i \end{array} \right ] + \frac{1}{\Delta x} \left [\begin{array}{c} \mathcal{A}_{\text{WENO-LF}} ({\bf u}_i) + \mathcal{H} ({\bf u}_i,C_i) \\ \mathcal{B}_{\text{WENO}} ({\bf u}_i, C_i) \end{array} \right ] = 0
\end{equation}
where $\mathcal{B}_{\text{WENO}}$ is given by \eqref{eqn:BWENO} and $\mathcal{A}_{\text{WENO-LF}}$ corresponds to the choice of the WENO flux described in \cite{Jiang1996202} (i.e. if $\mathcal{H} \equiv 0$ then \eqref{eqn:wenoLFCSemiDiscrete} is the same as WENO-LF). The term $\mathcal{H}$ is a discrete approximation of
$\tilde \beta \epsilon^2 C^{\epsilon,\delta} \rho^\epsilon |\partial_x u^\epsilon|^2$.
  The operator $\mathcal{H}$ is defined as
\[
\mathcal{H}({\bf u}_i,C_i) = \left [ \begin{array}{c} 0 \\ 0 \\ \beta \Delta x^2 \underset{i}{\max} | \tilde \partial u_{i+1/2}| \frac{C_i}{\underset{i}{\max} C_i} \rho_i |\tilde \partial u_i|^2 \end{array} \right ]. 
\]

In Figure \ref{subfig:LBLFbeta5Converg} we demonstrate the benefit of WENO-LF-C with $\beta = 5.0$, again at successive refinements of $N=360,720,1440$. The overshoot at the contact discontinuity is relatively non-existent while the shock speeds are far more accurate and appear to converge to the correct speed at a faster rate.

\section{Concluding Remarks}  \label{sec:conclude}
We have presented 
a localized space-time smooth artificial viscosity algorithm, the C-method, 
and have demonstrated its efficacy on a variety of classical
one-dimensional shock-tube problems. As compared to more
established procedures, the C-method has been shown to be 
highly
competitive with regards to  accuracy and stability, 
while being relatively easy to implement.
Because of its
simplicity, the C-method
can readily be extended to multiple space-dimensions and/or utilized in 
reactive-flow simulations. Of value to reactive flows is the
localized {\it smooth} diffusion provided by the C-method; specifically, the
function $C$  can be used to actively
influence various mixing-rate-limited reactions occurring near
sharp boundaries. 

In the future, the gradient-based source
term used in the current implementation of the C-method may be combined with   
a noise-indicator that turns off the current gradient-based source term when it is not needed.
Such noise-indicators require a very high-order scheme compatible with DG or 11th-order
WENO to name just two examples.  By projecting the solution onto a suitable basis,
the noise-indicator would activate when small-scale coefficients of this basis do not
have sufficient decay; in turn, an  indicator function, localized about the region of noise,
would activate and force the C equation.  This approach is taken in \cite{Barter20101810},
but without any gradient-based forcing functions like our function $G$ or $G_{comp}$.

For example, after the rapid initial growth
of the internal energy field in the Leblanc shock-tube problem, 
this field is essentially representative
of the advection of a square-wave. Thus, after  initial growth, the gradient-based
 source term in the C equation for energy 
could be deactivated leading to less diffusion in the downstream contact discontinuity;
simultaneously, the noise-indicator would
activate if small-scale instabilities were to set in.

 But, for more general
problems, the impact of the activation/deactivation of the
source term in the C-method on numerical accuracy is not entirely
obvious and is left for future research.

\section*{Acknowledgments}
SS  and JS were supported by the National Science Foundation under grant
DMS-1001850.  SS was partially supported by the United States Department of Energy through  Idaho National
Laboratory LDRD Project NE-156.
We thank Len Margolin for numerous discussions and helpful comments on early drafts of the manuscript. We are grateful to  Bill Rider for providing us data from his WENO-G 
scheme which we use in some of our comparisons.    We would also like to thank the anonymous referees for their comments and suggestions, which have both improved
and corrected our presentation.

\bibliographystyle{elsarticle-num}
\bibliography{myrefs}

\end{document}